\documentclass[]{elsart3}

\usepackage{amssymb}
\usepackage{graphicx}

\usepackage{epsfig}

\begin{document}

\begin{frontmatter}

\title{Non-Arrhenius relaxation of the Heisenberg model with dipolar and anisotropic interactions}

\author{Rogelio D\'\i az-M\'endez}
\address{Nanophysics Group, Department of Physics, Electric Engineering Faculty, CUJAE, ave 114 final, La Habana, Cuba}
\address{``Henri Poincar\'e'' Group of Complex Systems, Physics Faculty, University of Havana, La Habana, CP 10400, Cuba}

\author{Roberto Mulet}
\address{Department of Theoretical Physics, Physics Faculty, University of Havana, La Habana, CP 10400, Cuba}
\address{``Henri Poincar\'e'' Group of Complex Systems, Physics Faculty, University of Havana, La Habana, CP 10400, Cuba}

\date{May 2011}

\begin{abstract}

The dynamical properties of a 2D Heisenberg model with dipolar interactions and perpendicular anisotropy are studied using Monte Carlo simulations in two different ordered regions of the equilibrium phase diagram.  We find a temperature defining a dynamical transition below which the relaxation suddenly slows down and the system aparts from the typical Arrhenius relaxation to a Vogel-Fulcher-Tamann law. This anomalous behavior is observed in the scaling of the magnetic relaxation and may eventually lead to a freezing of the system.  Through the analysis of the domain structures we explain this behavior in terms of the domains dynamics. Moreover, we calculate the energy barriers distribution obtained from the data of the magnetic viscosity. Its shape supports our comprehension of both, the Vogel-Fulcher-Tamann dynamical slowing down and the freezing mechanism.  
\end{abstract}

\end{frontmatter}

\maketitle

\section{Introduction}

Thin magnetic films have been the subject of intense attention over the last two decades \cite{mac95,wu05,vater00}. These studies have been motivated both by the technological importance of these systems and by the insight that they provide into the fundamental role of interactions at the atomic level. Advances in film growth techniques and experimental control of spin-spin interactions have impulsed this interest. 

It is well established now the fundamental role of competing interactions in the emerging features of low dimensional systems. Among a wide scope of theoretical and numerical investigations on equilibrium\cite{mar07,lucas07,rm10} and dynamical properties\cite{glei02,rdm05,mulet07} of several model Hamiltonian of low-dimensional magnetic systems, Heisenberg-like models are one of the most intensively used to approach real magnetic materials.\cite{wang01,mar07}  In fact, anisotropic versions of the Heisenberg model describe several compounds as BaCo$_2$(AsO$_4$)$_2$,\cite{regnault77} CoCl$_2$-GIC\cite{wiesler87} and Rb$_2$CrCl$_4$\cite{castro02}

Some effort in the study of the non-equilibrium regime of closely related models has been done\cite{jang97,jang03,iglesias04,liu09} and a detailed characterization of the equilibrium features of the  Heisenberg Hamiltonian considering a dipolar long-range repulsive term and a varying perpendicular anisotropy  were depicted by numerical and analytical methods in the last few years.\cite{mar07,sant09,sant10} 
In particular, the  transition in which spins reorient from an in-plane ferromagnetic order to an out-plane perpendicular stripes with decreasing temperature, already seen in experimental works,\cite{wong05} was encountered in a well-localized anisotropy range. However, a complete characterization of this system is far from been established, and that is particularly true   in relation with its dynamical behavior. In this paper we want to make a step forward in this direction.

Through Monte Carlo simulations we study the dynamics of the model analyzing the magnetic relaxation behavior  for several values of temperature. We explore dynamical properties in two different ordered regions of the phase space near the Spin Reorientation Transition.
We show that there is a temperature  $T^*$ below which the dynamic slows down.
This behavior is present for the two phases around the reorientation transition.
We explain this  behavior in terms of non-Arrhenius relaxation due to complex domain dynamics. 

The paper is organized as follows: In section \ref{mss} we present the model making a review of some of its more relevant properties and give details about the Monte Carlo simulations. Then, in section \ref{rd} we present and discuss our results understanding the anomalous behavior of magnetic relaxation by means of the analysis of the domains dynamics and the energy barriers associated to the different regions.
 Finally, section \ref{conc} is devoted to summarize our results and present the conclusions of the work.

\section{Model and simulation}
\label{mss}

We consider a two-dimensional square lattice of Heisenberg spins 
of unit modulus $|\vec{s}_i|=1$ with an anisotropy  perpendicular to the plane of the lattice, interacting through the dimensionless Hamiltonian

\begin{eqnarray}
\nonumber
H &=& -\eta\sum_i (s_i^z)^2-\delta\sum_{\langle ij\rangle }\vec{s}_i\cdot\vec{s}_j+\\
&+&\sum_{i\neq j}\left( \frac{\vec{s}_i\cdot\vec{s}_j}{r_{ij}^3}-3\frac{(\vec{s}_i\cdot\vec{r}_{ij})(\vec{s}_j\cdot\vec{r}_{ij})}{r_{ij}^5}\right) 
\label{ham}
\end{eqnarray}

\noindent where the exchange $\delta$ and anisotropy constants $\eta$ 
are normalized relative to the dipolar coupling constant, $< i, j >$ stands for
a sum over nearest neighbors pairs of sites in the lattice, (i, j)
stands for a sum over all distinct pairs and $r_{ij} = |\vec{r}_i-\vec{r}_j |$ is
the distance between spins $i$ and $j$. 

Recently, Carubelli et all\cite{mar07}  determined the $\eta$-$T$ phase diagram of this model by means of Monte Carlo simulations. Maintaining the same relative strength between exchange and dipolar interactions they showed the existence of two different ordered phases at low temperature. A planar ferromagnet or a perpendicular striped phase establishes depending on the value of $\eta$.
Moreover, the diagram obtained in \cite{mar07} qualitatively reproduces detailed measurements of the reorientation transition encountered experimentally in samples of Fe/Ni/Cu(001)\cite{wong05} suggesting that the film thickness acts as an effective inverse anisotropy.

\begin{figure}[!htb]
\includegraphics[angle=-90,width=.45\textwidth]{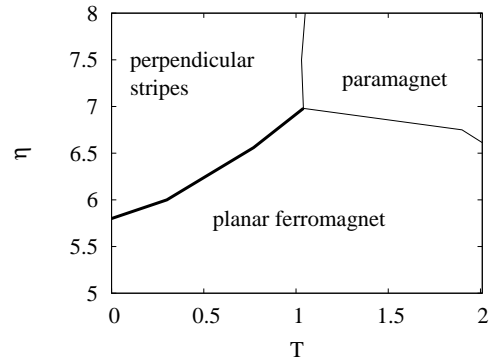}
\caption{Scheme of the phase diagram of a system of size $32\times32$ for $\delta=3$. The dynamical calculations are done near the Spin Reorientation Transition (shown as a bold line).}
\label{pd} 
\end{figure}

We focused on the study of quenches for a system of size $N=48\times48$ and $\delta=3$ from full out-of-plane magnetized configurations into planar and striped ordered regions. 
Several temperatures below $T=0.9$ corresponding to the ordered phases were explored for two anisotropy values $\eta=5.5$ and $\eta=7.0$.
In figure \ref{pd} a scheme of the phase diagram is shown and the Spin Reorientation Transition is marked with a bold line. 

All the simulations were done using the Metropolis algorithm and infinite periodic boundary conditions were imposed on the lattice by means of the Ewald sums technique. The trial for the spin flip was chosen as a random orientation uniformly distributed in the sphere.

The time evolution of the out-of-plane magnetization $M_z=1/N\ \sum_is_i^z$ was recorded starting from the saturated configuration $s_i^z=1$. 
The behavior of this observable is in all cases a monotonous decay to zero due to the lack of total magnetization of the striped configurations as well as the complete in-plane orientation of spins in the planar-ferromagnetic phase. Typical time orders in Monte Carlo steps were $t=10^5$ m.c.s. and averages over thermal noise involved a few hundred  of realizations. In the quench to the striped phase $\eta=7.0$, it was impossible to relax the system at temperatures below $T=0.4$. There was not clear  magnetization decay up to $t=10^7$ m.c.s.

\section{Results and Discussion}
\label{rd}

\begin{figure}[!thb]
\includegraphics[angle=-90,width=.45\textwidth]{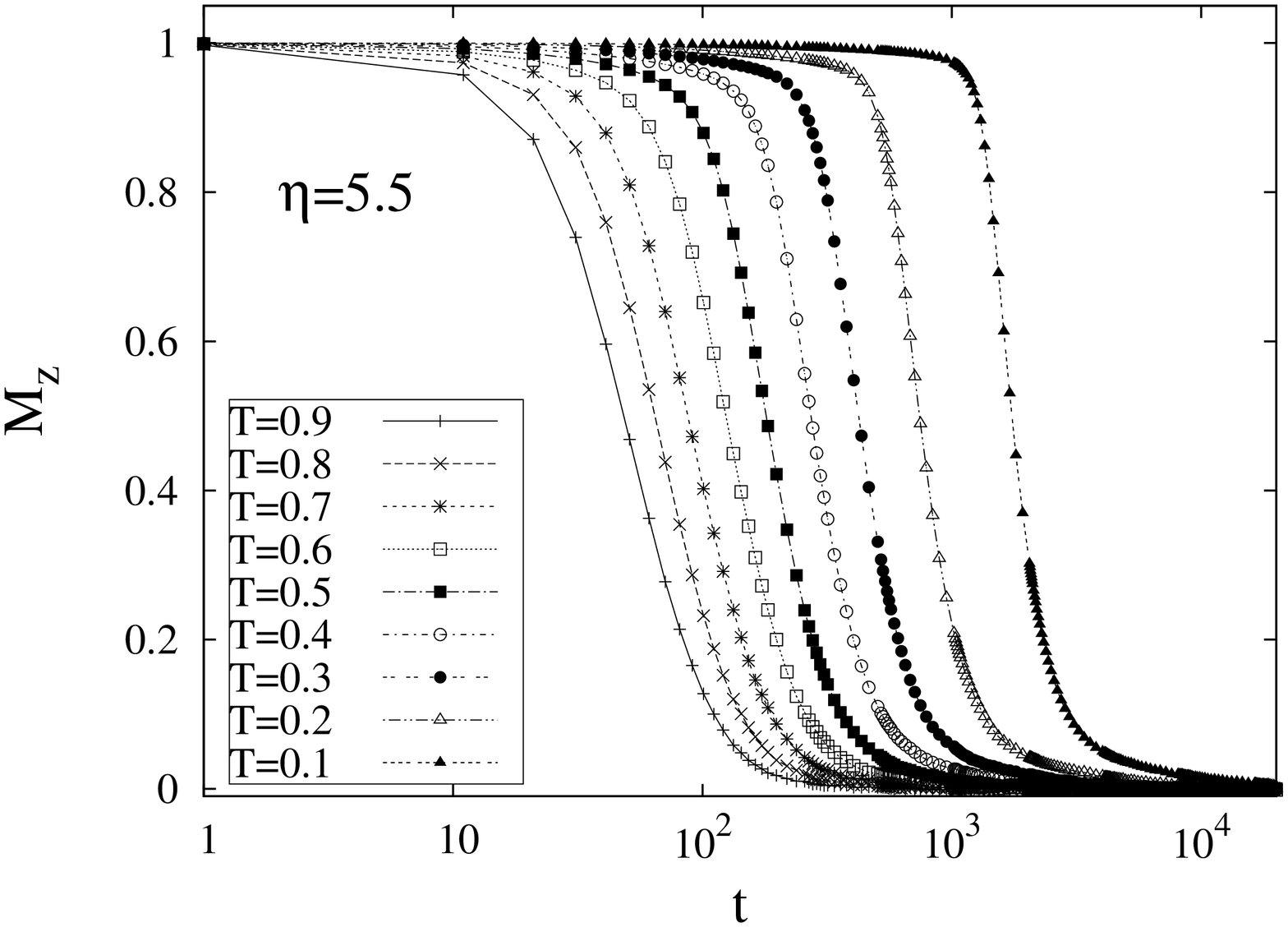}\\
\includegraphics[angle=-90,width=.45\textwidth]{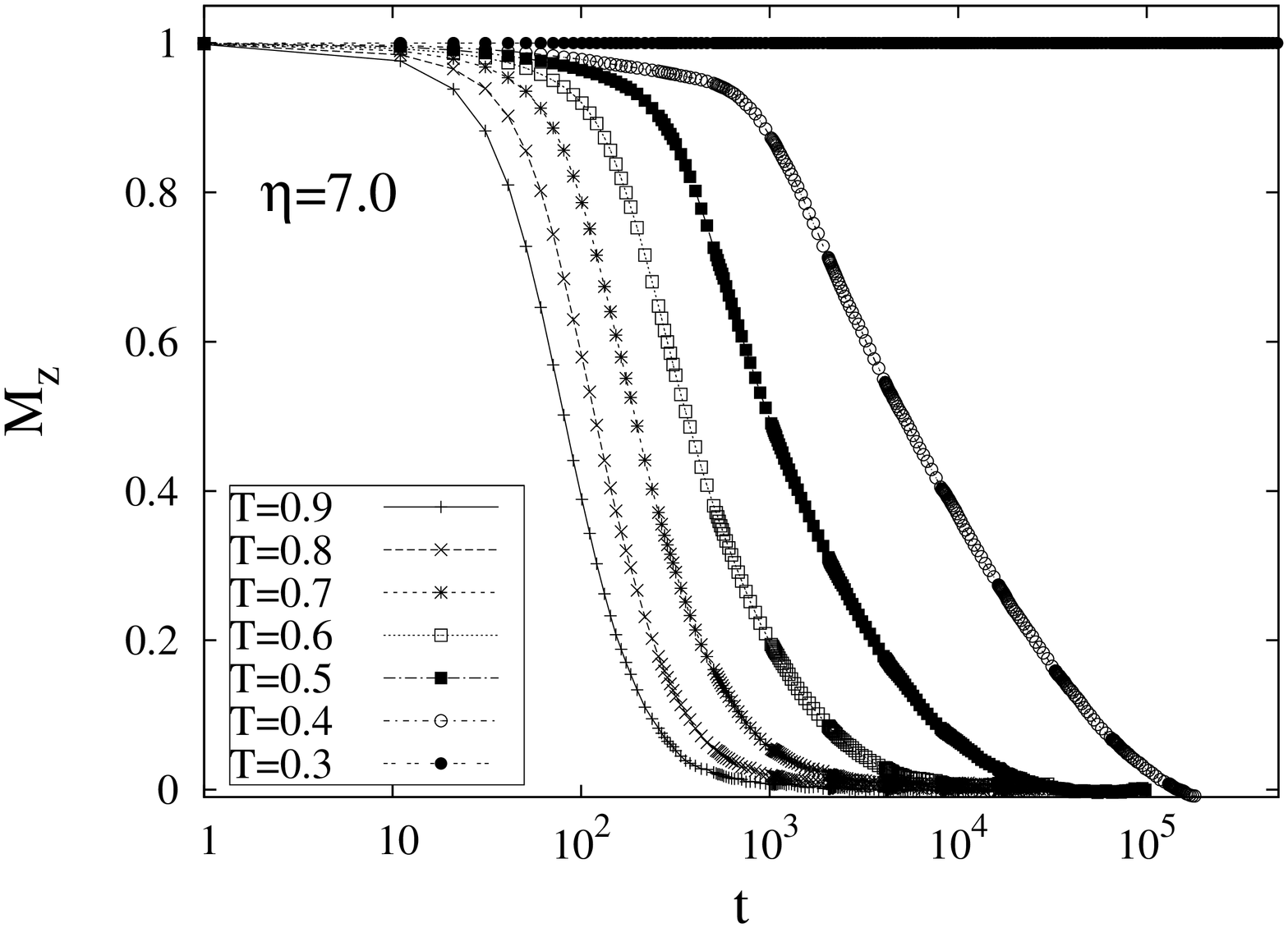}
\caption{Out-plane magnetization for several temperatures with $\eta=5.5$ (up) and $\eta=7.0$ (down).}
\label{ms} 
\end{figure}

We start showing in figure \ref{ms} the out-of-plane magnetization relaxation corresponding to both quenches at different temperatures. 

As can be seen there is a central difference between quenches in the two ordered regions.  For $\eta=7.0$ the ground state of the system is a striped configuration and, as we already mentioned, there exists a temperature $T=0.4$ below which the dynamics completely freezes.
Contrary to the $\eta=5.5$ case where degeneration of planar ferromagnetic ground state facilitates domain relaxation, for  $\eta=7.0$ the dynamical flipping of spins corresponding to adjacent striped domains is naturally slower. 

\begin{figure}[!htb]
\includegraphics[angle=-90,width=.45\textwidth]{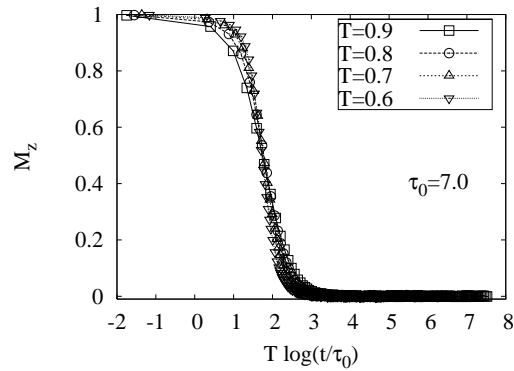}
\caption{Scaling of magnetization for several temperatures with $\eta=5.5$.}
\label{Msc5.5} 
\end{figure}

The scaling of $M_z$ for $\eta=5.5$ at high temperatures is shown in figure \ref{Msc5.5} as a function of $T\mathrm{ln}(t/\tau_0)$.  There was no possible scaling in this functionality for lower temperatures.

It means that below  $T\leq T^*=0.5$ the system does not longer relax following the Arrhenius law. \cite{street49} In fact, despite many works validating the applicability of the scaling of figure \ref{Msc5.5} to interacting systems,\cite{iglesias02,iglesias04} there is no guarantee that interacting spins may be always described in terms of thermal activation of the Arrhenius type.

The most simple and common extension for the Arrhenius equation is the well known Vogel-Fulcher-Tammann (VFT) law which is given by 
\begin{equation}
\tau(E)=\tau_0  e^{\frac{U}{T-T_0}}
\label{vftl}
\end{equation}
where the new parameter $T_0$ is known as the VFT temperature. This functionality provides usually a good description of experimental and numerical data and is the salient phenomenological description for the $\alpha$-relaxation dynamics of glass-forming systems.\cite{angell00} 

Although the microscopic origin of the VFT law is still under debate, it is generally accepted that a clue feature of it is the notion of cooperativity in the movement of the elementary constituents of the system.
Thus, many theories of the glass transition are based on the assumption that cooperatively rearranging regions (domains or clusters) merges at low temperatures. Inside them magnetic moments move in some concerted way that distinguishes these regions from their surroundings.\cite{rault00,langer07}

\begin{figure}[!htb]
\includegraphics[angle=-90,width=.45\textwidth]{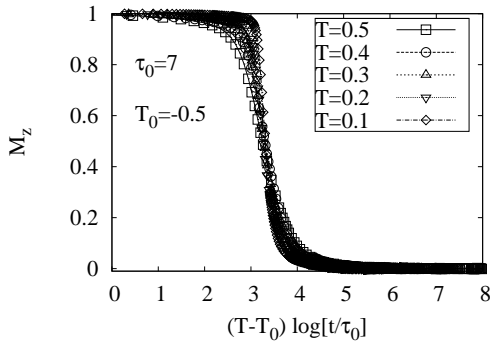}
\caption{Scaling of magnetization for all temperatures at $\eta=5.5$ considering a VFT relaxation law. Collapse is obtained for $\tau_0=7$ and $T_0=-0.5$.}
\label{Mscna} 
\end{figure}

From expression (\ref{vftl}) it is direct to see that the expected scaling variable now takes the form $(T-T_0)\ln(t/\tau_0)$.  In figure \ref{Mscna} this scaling is shown for temperatures below $T=.5$. 
On the contrary, there is no collapse for higher temperatures where the data follows the typical Arrhenius relaxation seen in figure \ref{Msc5.5}. 
It is worth to note that the value of the parameters for the best collapse are shown in the figure and involve a negative VFT temperature $T_0=-0.5$. As far as we know, this feature has being observed only in a few materials\cite{sch93,lunk96} and its interpretation remains unclear.  

\begin{figure}[!htb]
\includegraphics[angle=-90,width=.45\textwidth]{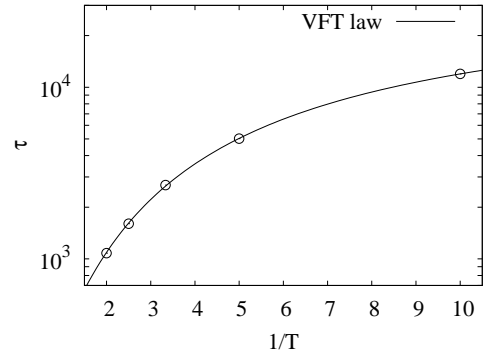}
\caption{Relaxation times of the sample magnetization for temperatures inside the non-Arrhenius regime. The solid line corresponds to the best fit with the VFT law (\ref{vftl}). Parameter values are $\tau_0=35\pm8$, $T_0=-0.47\pm0.03$ and $U=3.3\pm0.3$.}
\label{arrp} 
\end{figure} 

In order to confirm this unusual feature we studied  the relaxation times $\tau(T)$ defined as the time at which the out-plane magnetization decays to a value $M_z=0.01$. In figure \ref{arrp} we see the Arrhenius plot of $\tau$ and its fit using the VFT form. As can be noticed there is a very good agreement between the data extracted from the simulation and the expression (\ref{vftl}). Confirming our previous results, the VFT temperature extracted from this fitting is $T_0=-0.47\pm0.03$.

\begin{figure}[!htb]
\includegraphics[angle=-90,width=.45\textwidth]{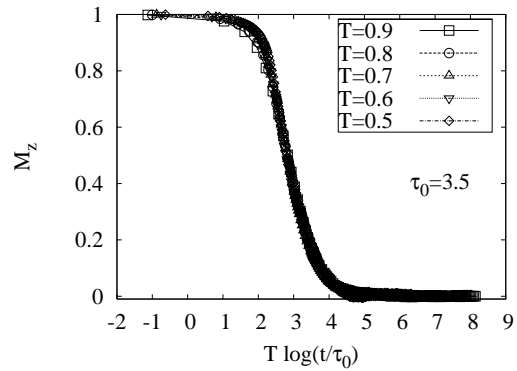}
\caption{Scaling of magnetization for several temperatures with $\eta=7.0$.}
\label{Msc7} 
\end{figure} 

For the $\eta=7.0$ case figure \ref{Msc7} shows that the Arrhenius-like relaxation holds down to a temperature $T^*=0.4$. The data corresponding to $T=0.4$ was impossible to collapse in the same figure and, since it is the only relaxation data below the Arrhenius zone, a proper collapse considering expression (\ref{vftl}) can not be found. The freezing of the system is stronger that for the case $\eta=5.5$.

\subsection{Cluster Analysis}
\label{ca}

Now it is worth to do a more deep study to understand the nature of the slow-down encountered that makes the system to apart from the Arrhenius law and  may eventually freeze the relaxation dynamics.
Firstly, the disagreement with the logarithmic scaling and better adjusting to a VFT relation reflects an intrinsic physical phenomenology of the model under study. That is, the emergence of larger times for the cooperative relaxations as temperature is decreased below $T^*$.  
Moreover, the dynamical freezing for $\eta=7.0$ could be a feature coming from the combined action of the slow non-Arrhenius relaxation and the context of a ground state of perpendicular-oriented spins that naturally develops higher relaxation times due to higher energy barriers between local minima.

\begin{figure}[!htb]
\begin{center}
\includegraphics[angle=-90,width=.23\textwidth]{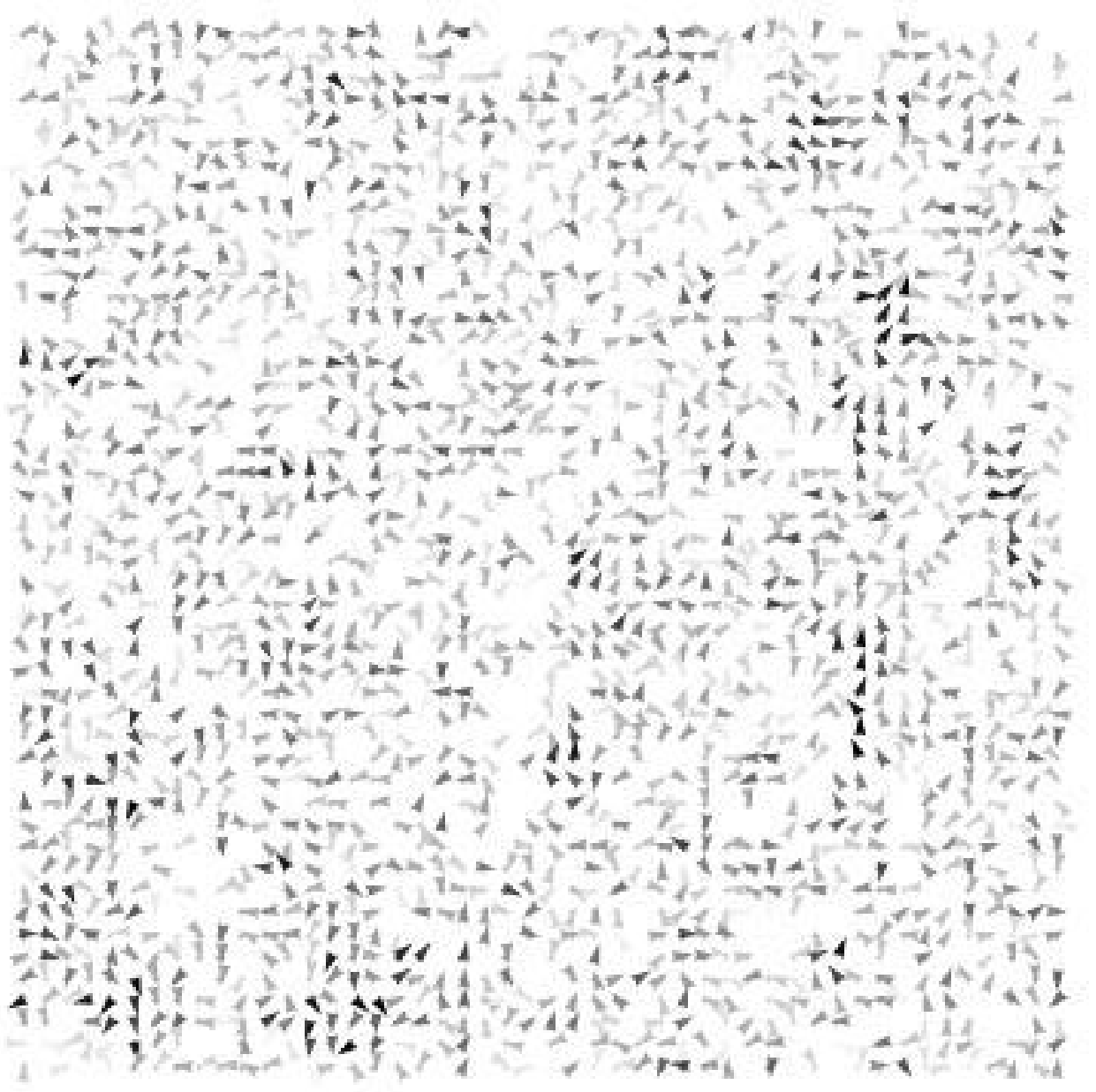}
\includegraphics[angle=-90,width=.23\textwidth]{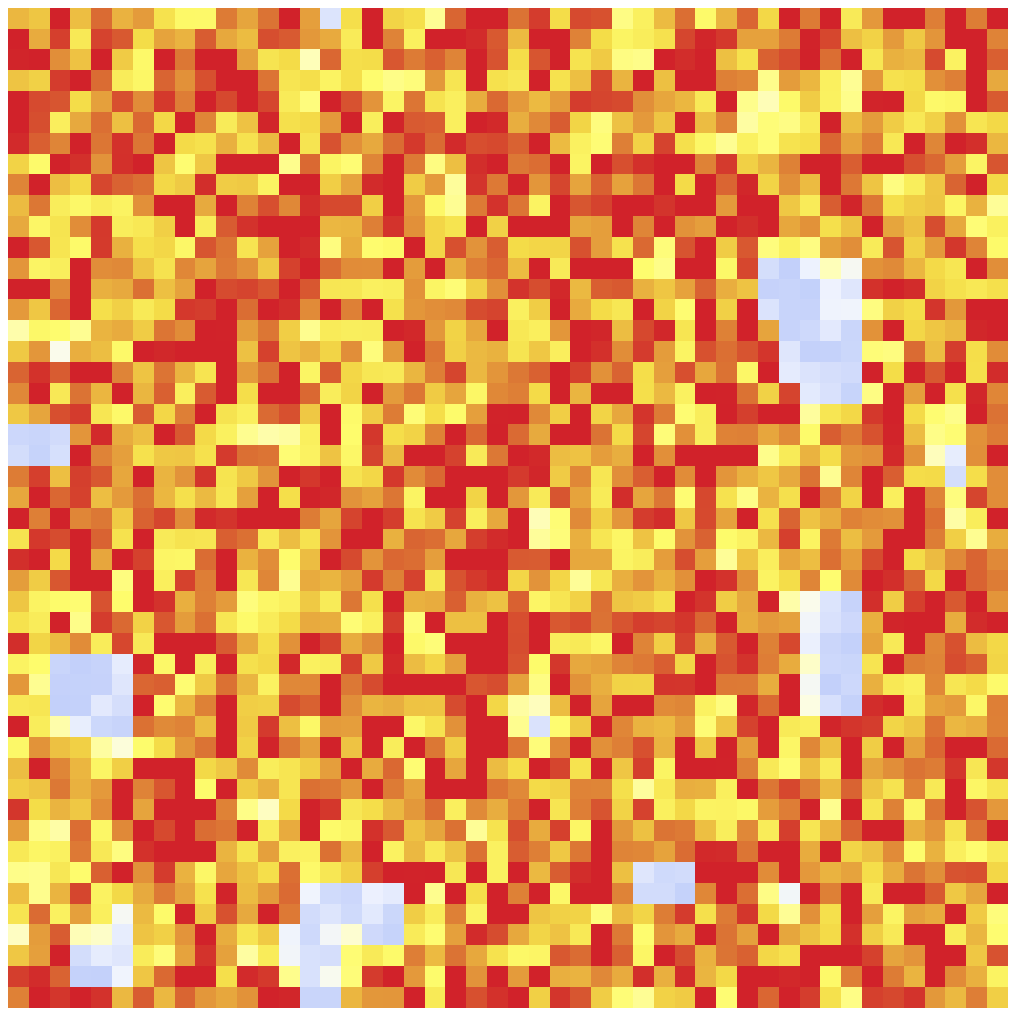}\\
\includegraphics[angle=-90,width=.23\textwidth]{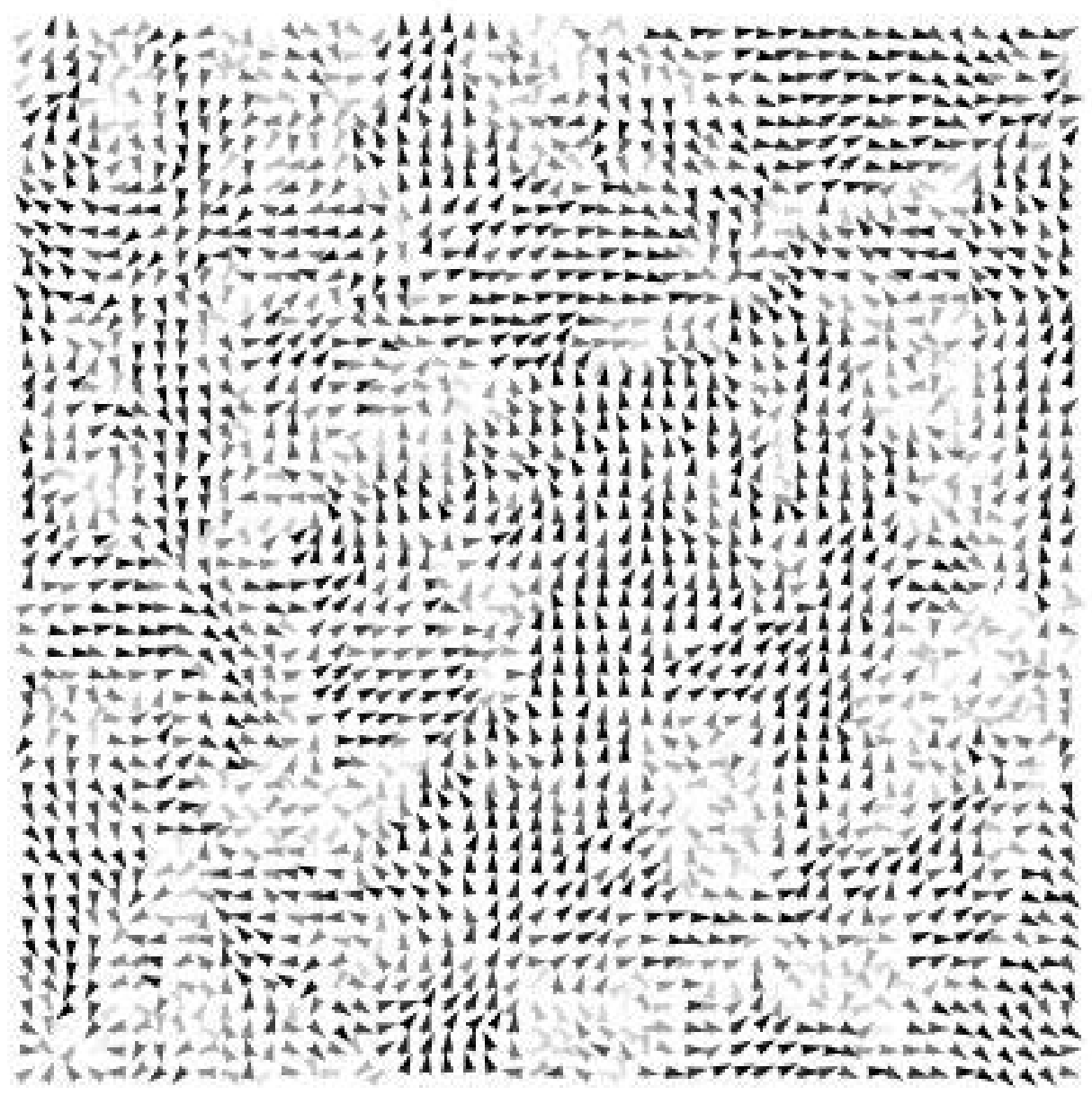}
\includegraphics[angle=-90,width=.23\textwidth]{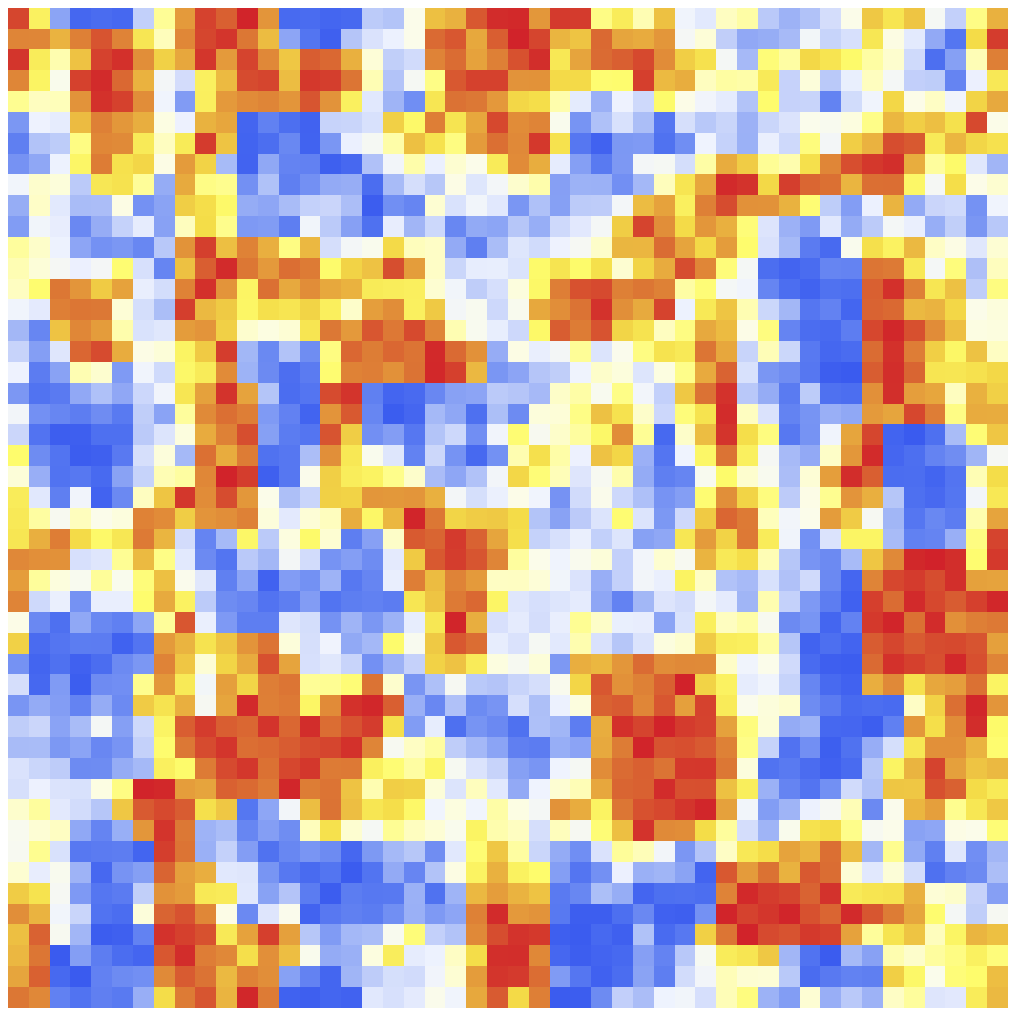}\\
\includegraphics[angle=-90,width=.23\textwidth]{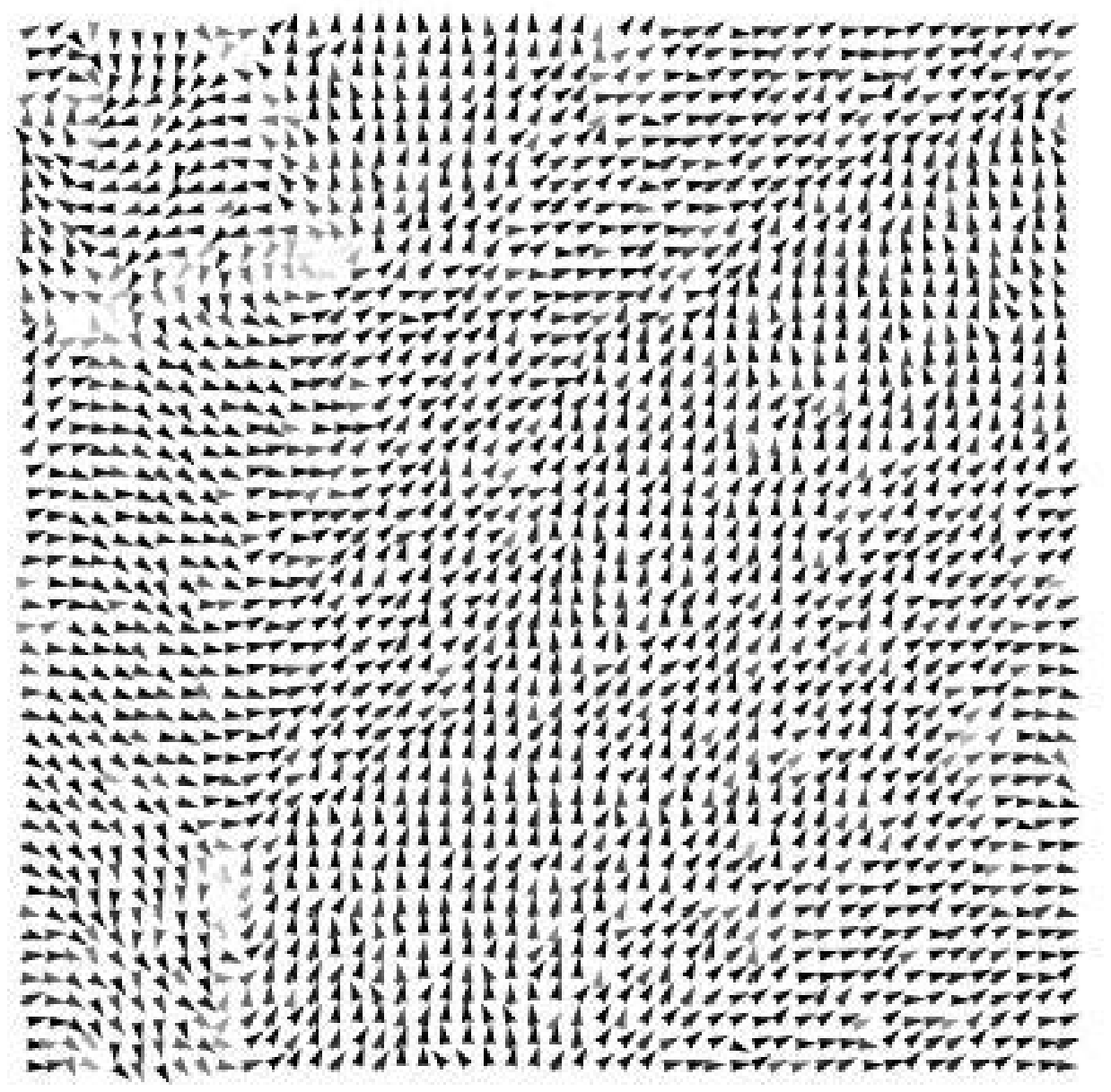}
\includegraphics[angle=-90,width=.23\textwidth]{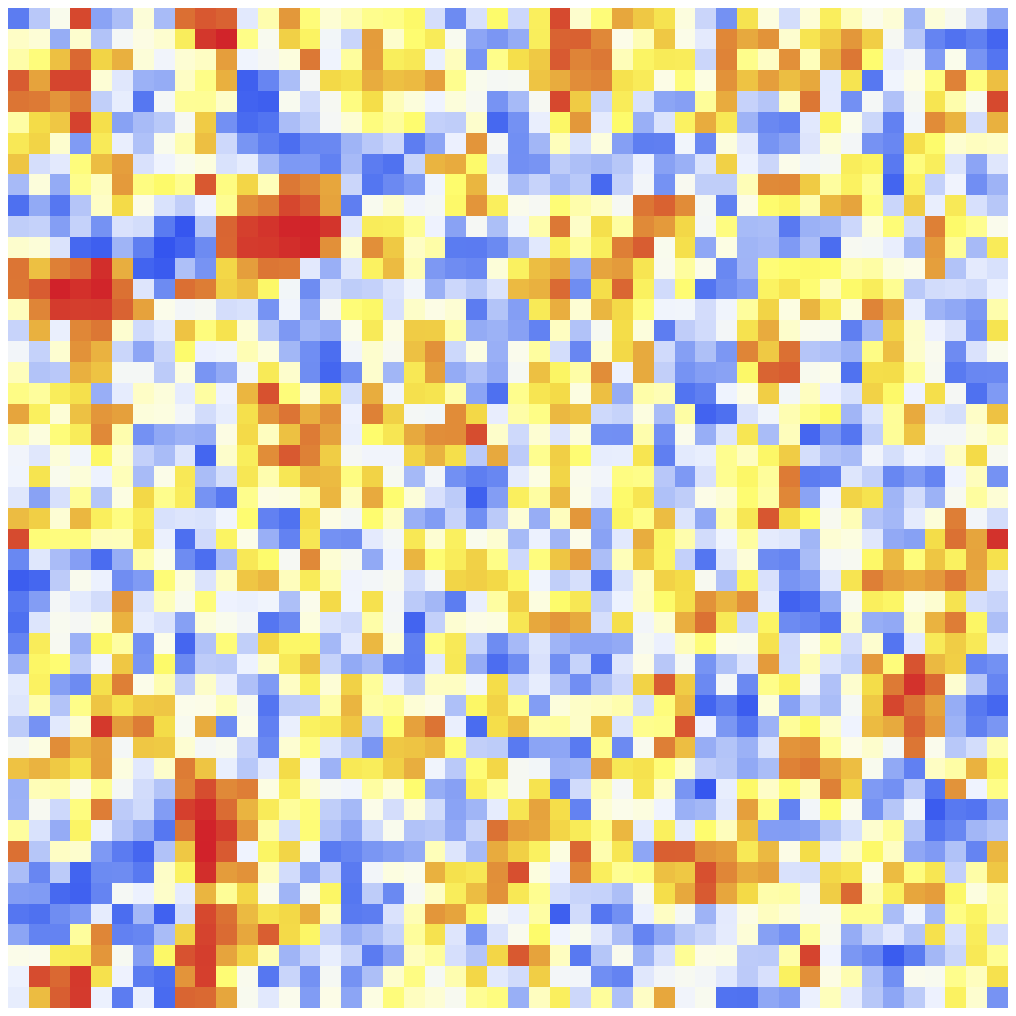}\\
\includegraphics[angle=-90,width=.23\textwidth]{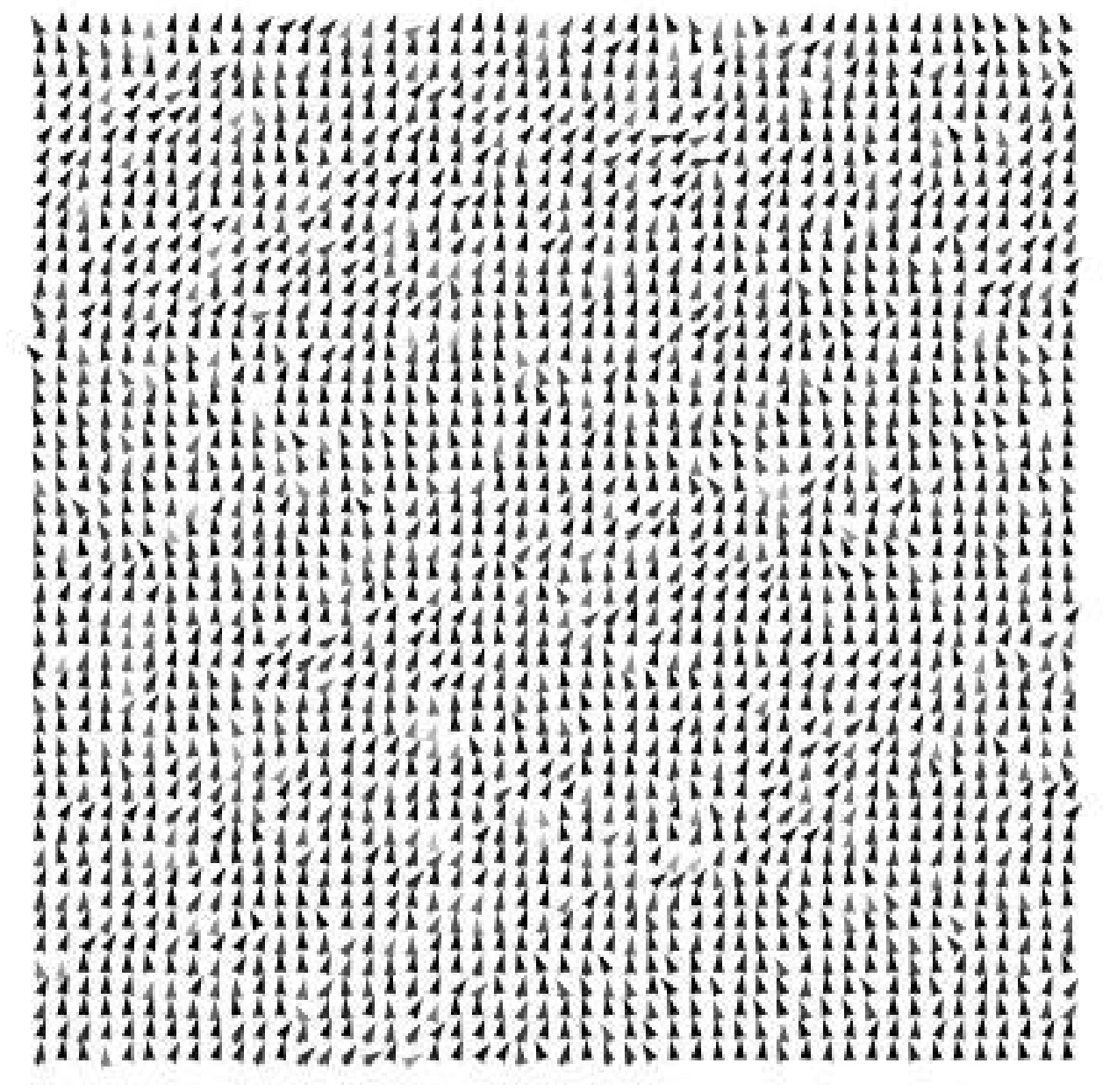}
\includegraphics[angle=-90,width=.23\textwidth]{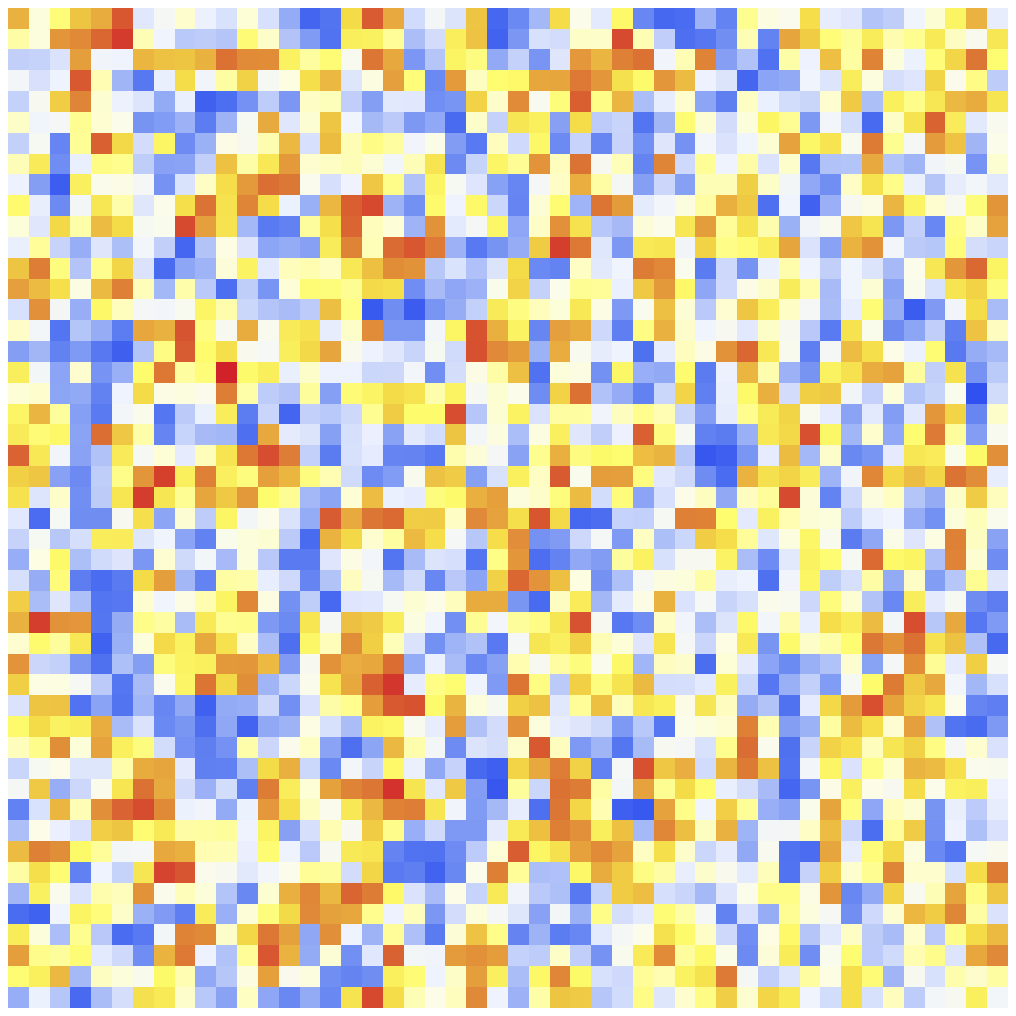}
\caption{Typical system configurations for the $\eta=5.5$ quench at $T=0.6$ and several times. The left (right) side shows the in-plane (out-plane) projection of the spins. From up to down correspondent times are 
$t=64, 256, 1024, 16385$.}
\label{confT.6} 
\end{center}
\end{figure}

Due to the three-dimensional nature of the spins in the system, the visual inspection of the configurations involves two type of formats: the in-plane and the out-plane projection. For the in-plane projection of the spins we use the format shown in the left side of figure \ref{confT.6}. There, the spin is represented with a single arrow and the dark intensity is proportional to the modulus of the projection, so that white corresponds to a completely out-plane spin. On the other hand, for the out-plane projection the format is seen in the right side of the same figure. In this case every spin is a tiny square and the modulus of the out-plane projection  goes from $-1$ (blue) to $1$ (red), so that white corresponds to a completely in-plane spin.

Figure \ref{confT.6} shows a typical evolution of the system with $\eta=5.5$ and $T=0.6$. Starting from the fully out-plane magnetized state, the spins begin to depart from this position, some of them even inverting its original direction. At some point several ground-state in-plane domains merges, these domains are eventually separated by out-plane structures that continuously disappears as the time grows and equals $\tau$ and the system orders as a planar ferromagnet.   

In order to compare this evolution for several temperatures we have to be careful. Since even in the Arrhenius regime ($T>T^*$) it is clear that the relaxation slows with decreasing temperature, direct inspections to complete evolutions above and below $T^*$ does not leads to useful conclusions. Instead, what we do is to fix certain equivalent stage, $M_z=0.5$, in the relaxation process at different temperatures and then inspect the typical configurations. 

\begin{figure}[!thb]
\begin{center}
\includegraphics[angle=-90,width=.23\textwidth]{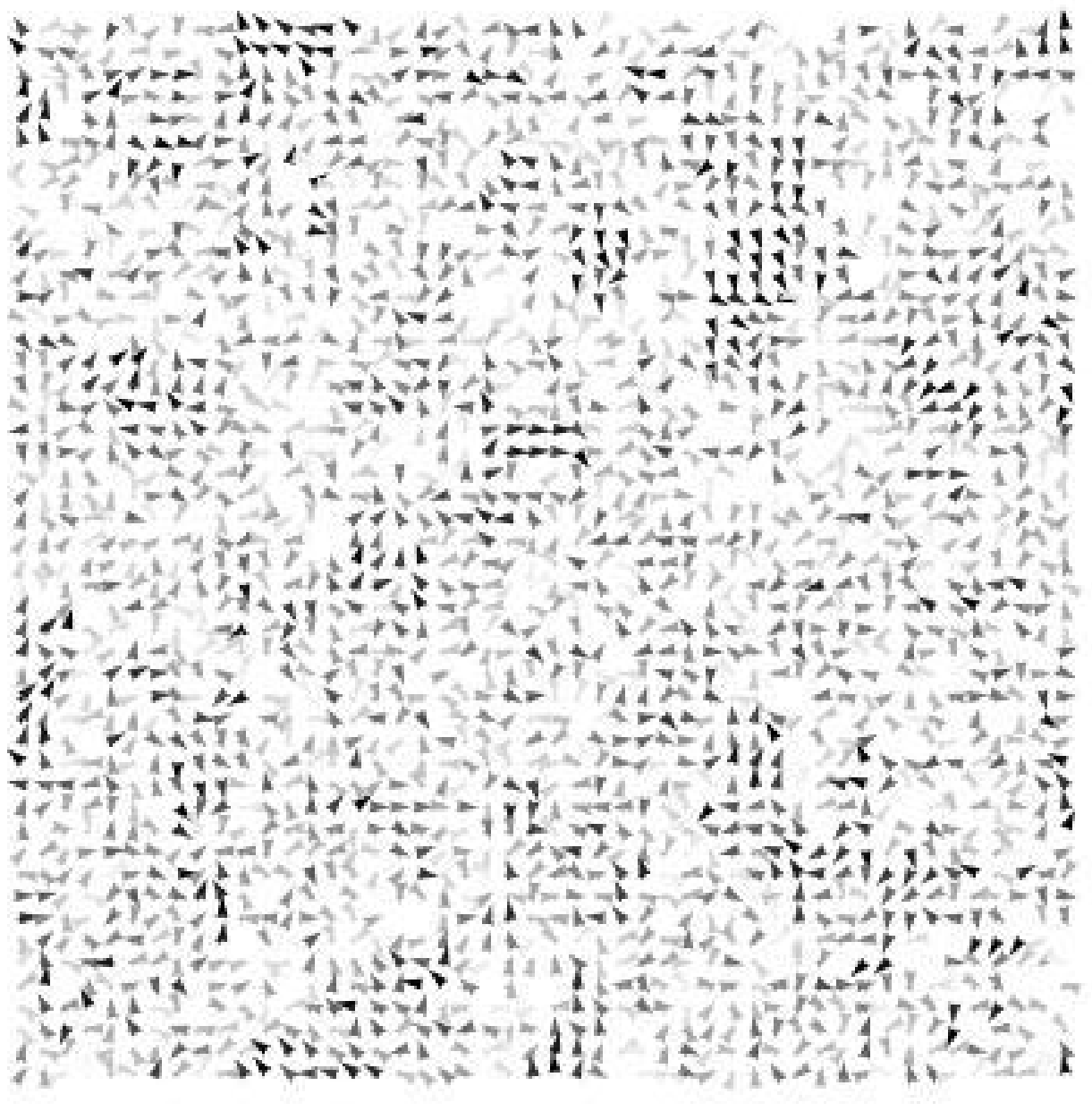}
\includegraphics[angle=-90,width=.23\textwidth]{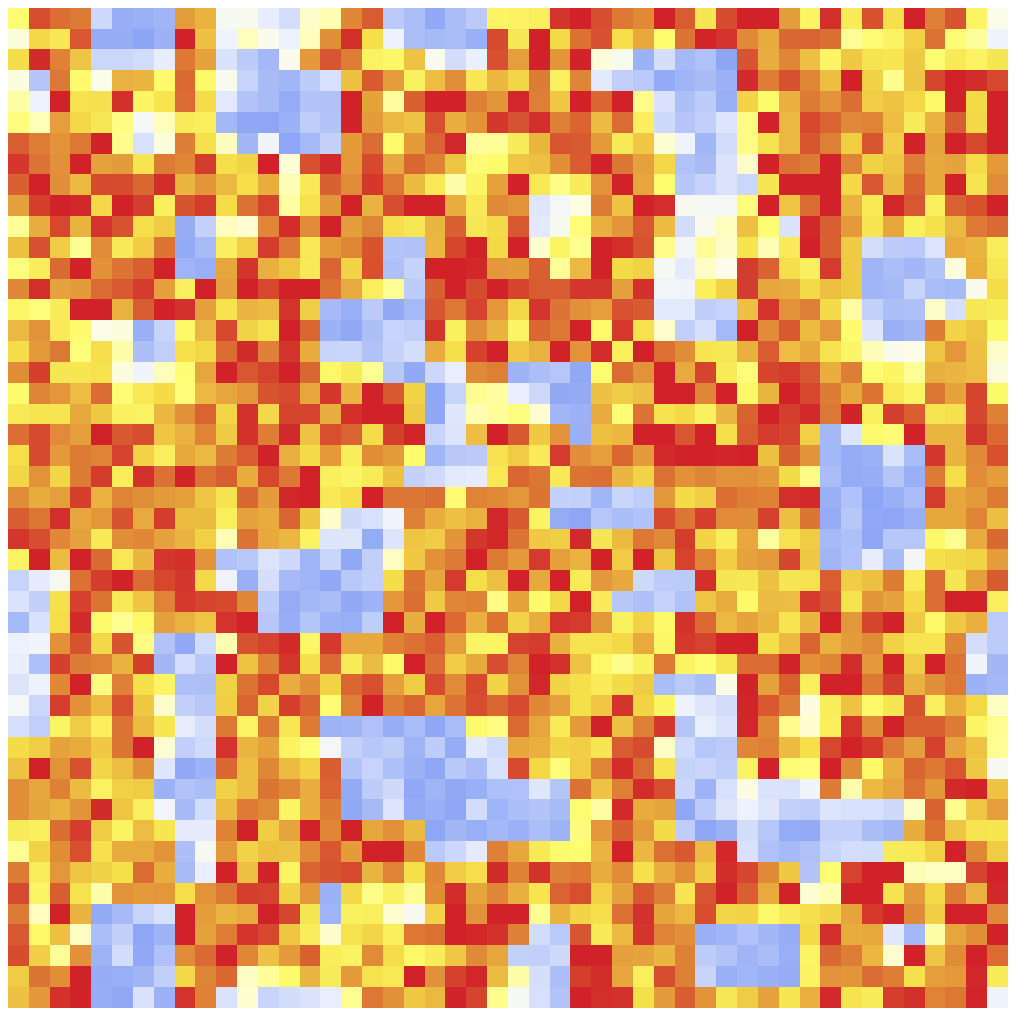}\\
\includegraphics[angle=-90,width=.23\textwidth]{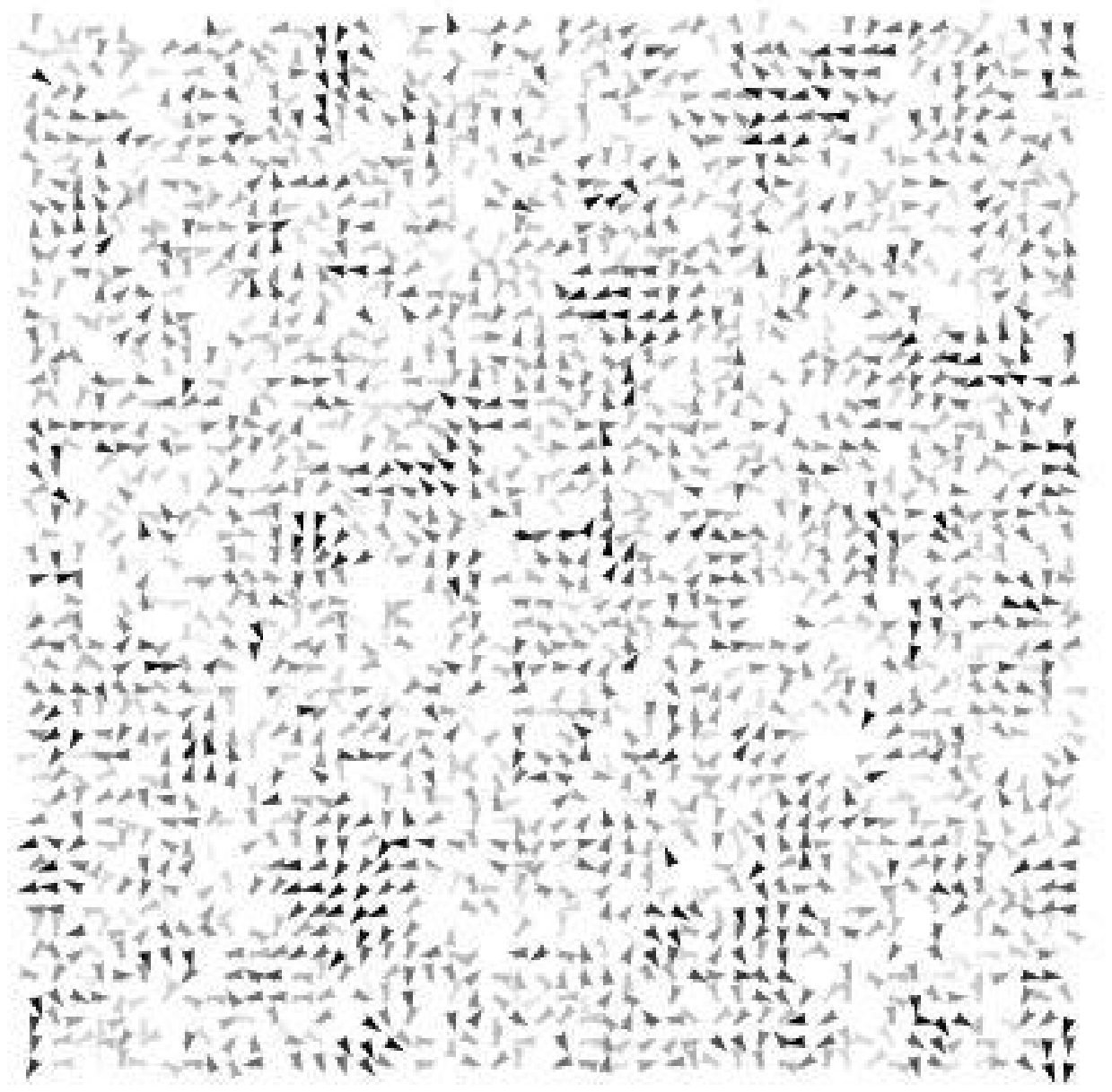}
\includegraphics[angle=-90,width=.23\textwidth]{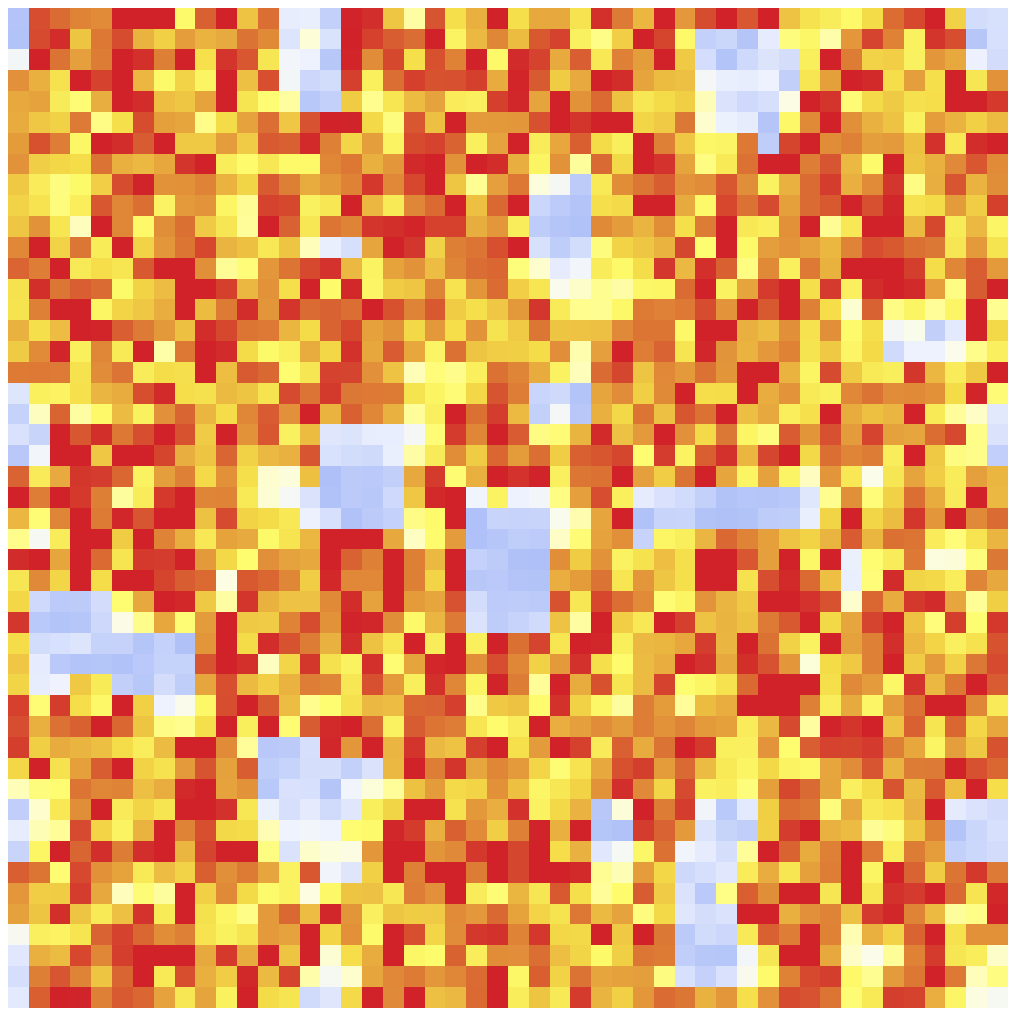}\\
\includegraphics[angle=-90,width=.23\textwidth]{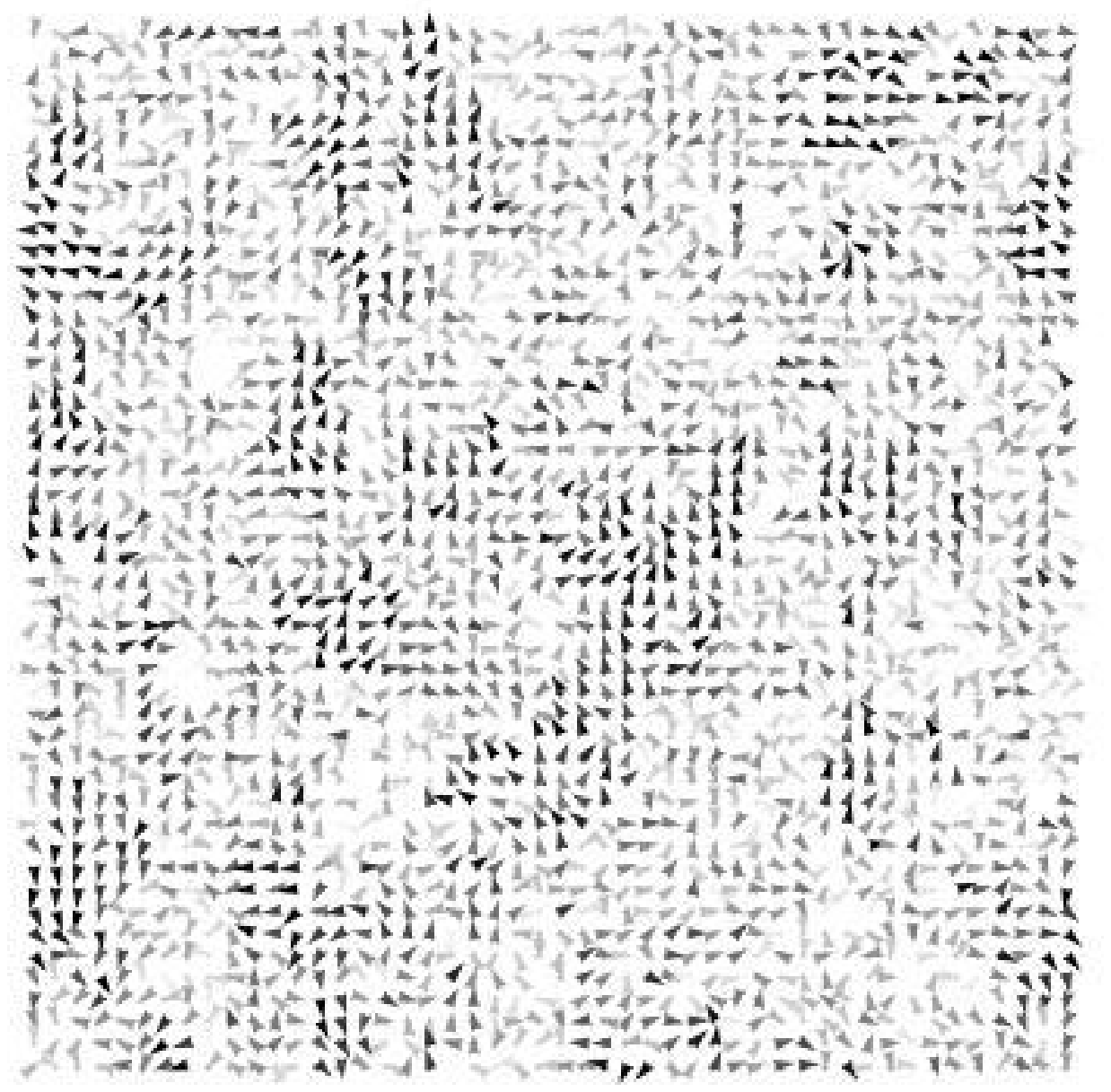}
\includegraphics[angle=-90,width=.23\textwidth]{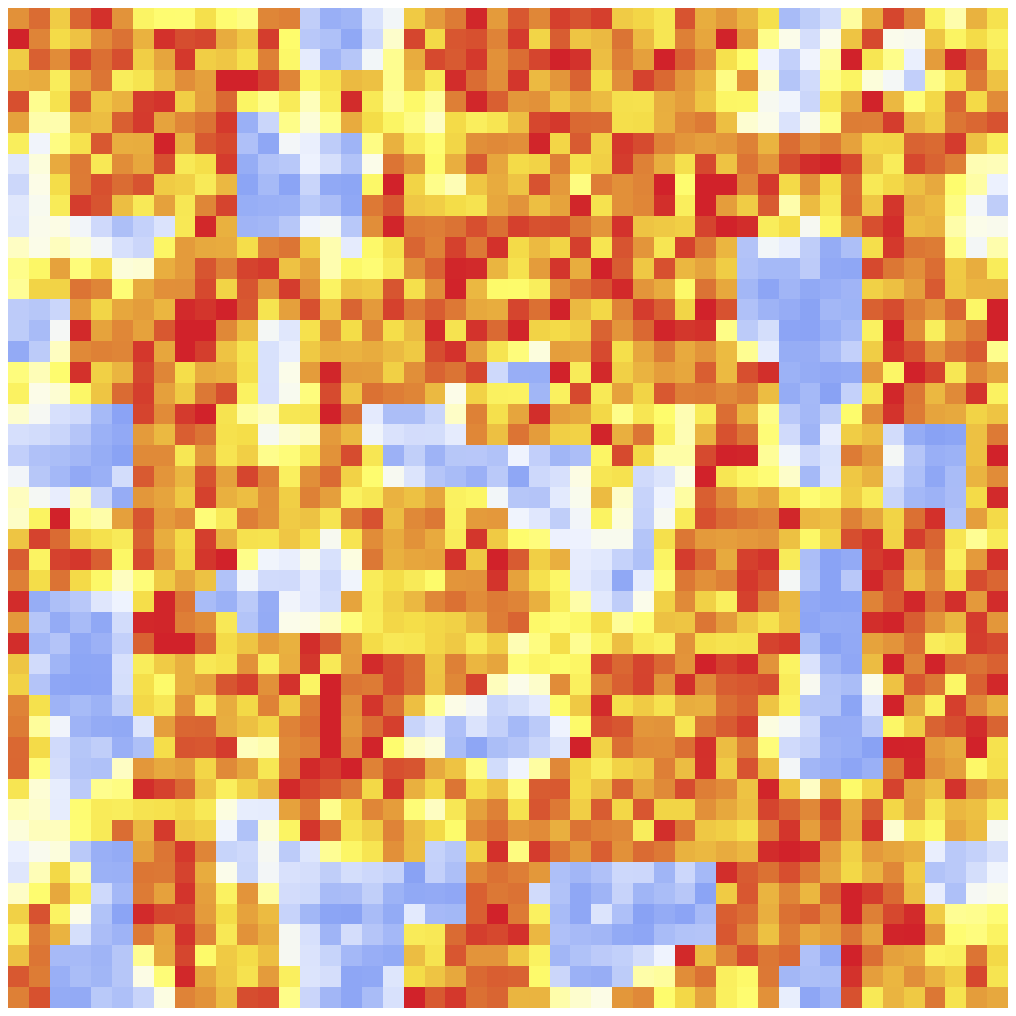}\\
\includegraphics[angle=-90,width=.23\textwidth]{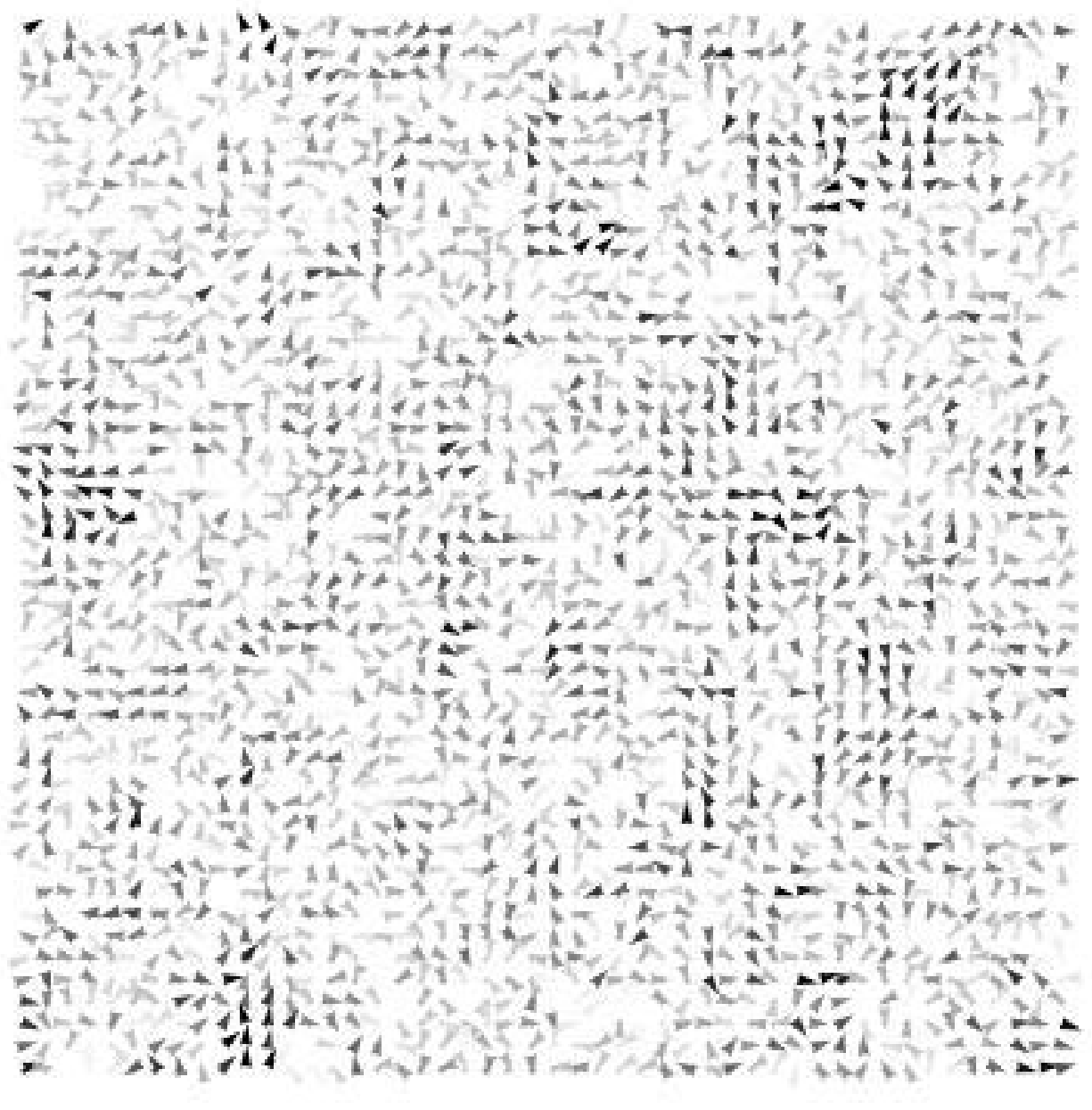}
\includegraphics[angle=-90,width=.23\textwidth]{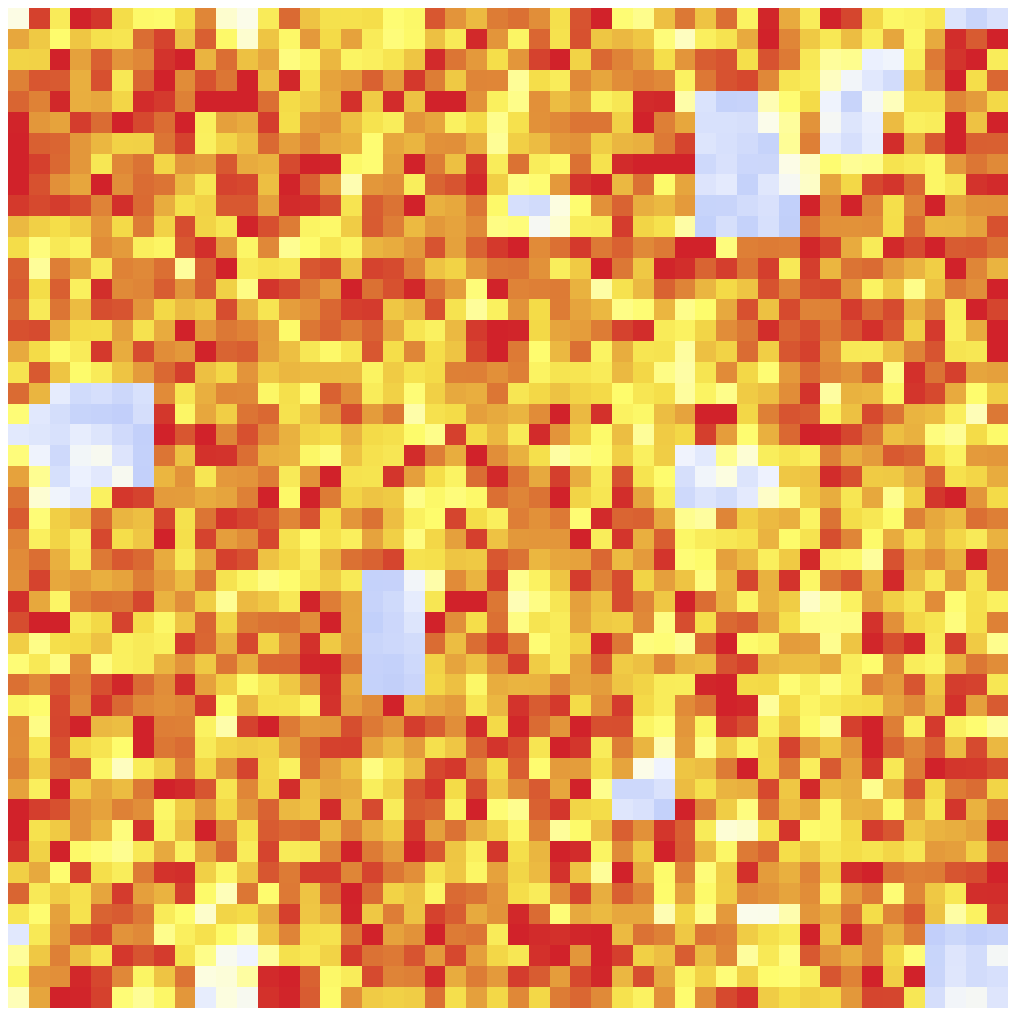}
\caption{Typical system configurations for the $\eta=5.5$ quench at $M_z=0.5$ and several temperatures above $T^*$. From up to down $T=0.8, 0.7, 0.6, 0.5$.}
\label{inconfMz.5} 
\end{center}
\end{figure} 

\begin{figure}[!thb]
\begin{center}
\includegraphics[angle=-90,width=.23\textwidth]{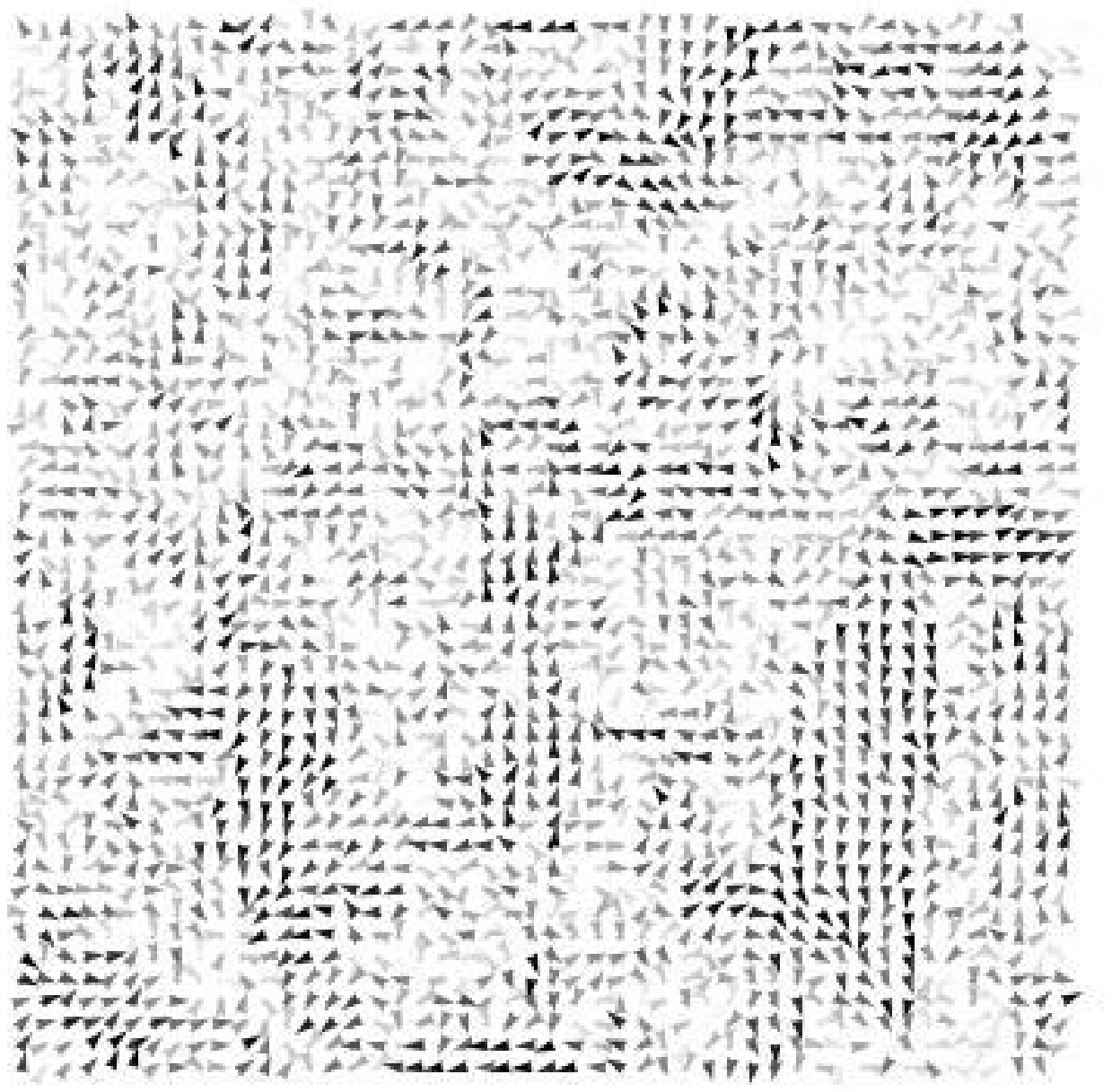}
\includegraphics[angle=-90,width=.23\textwidth]{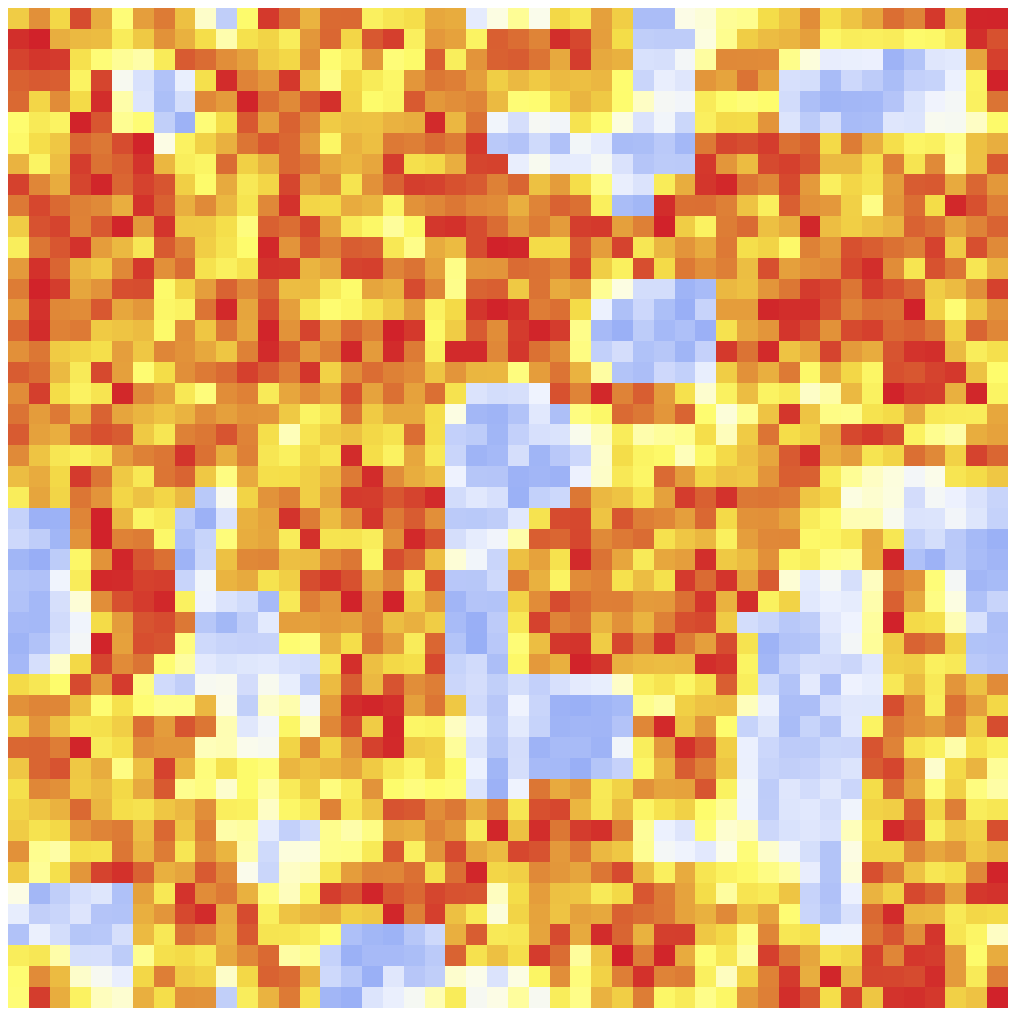}\\
\includegraphics[angle=-90,width=.23\textwidth]{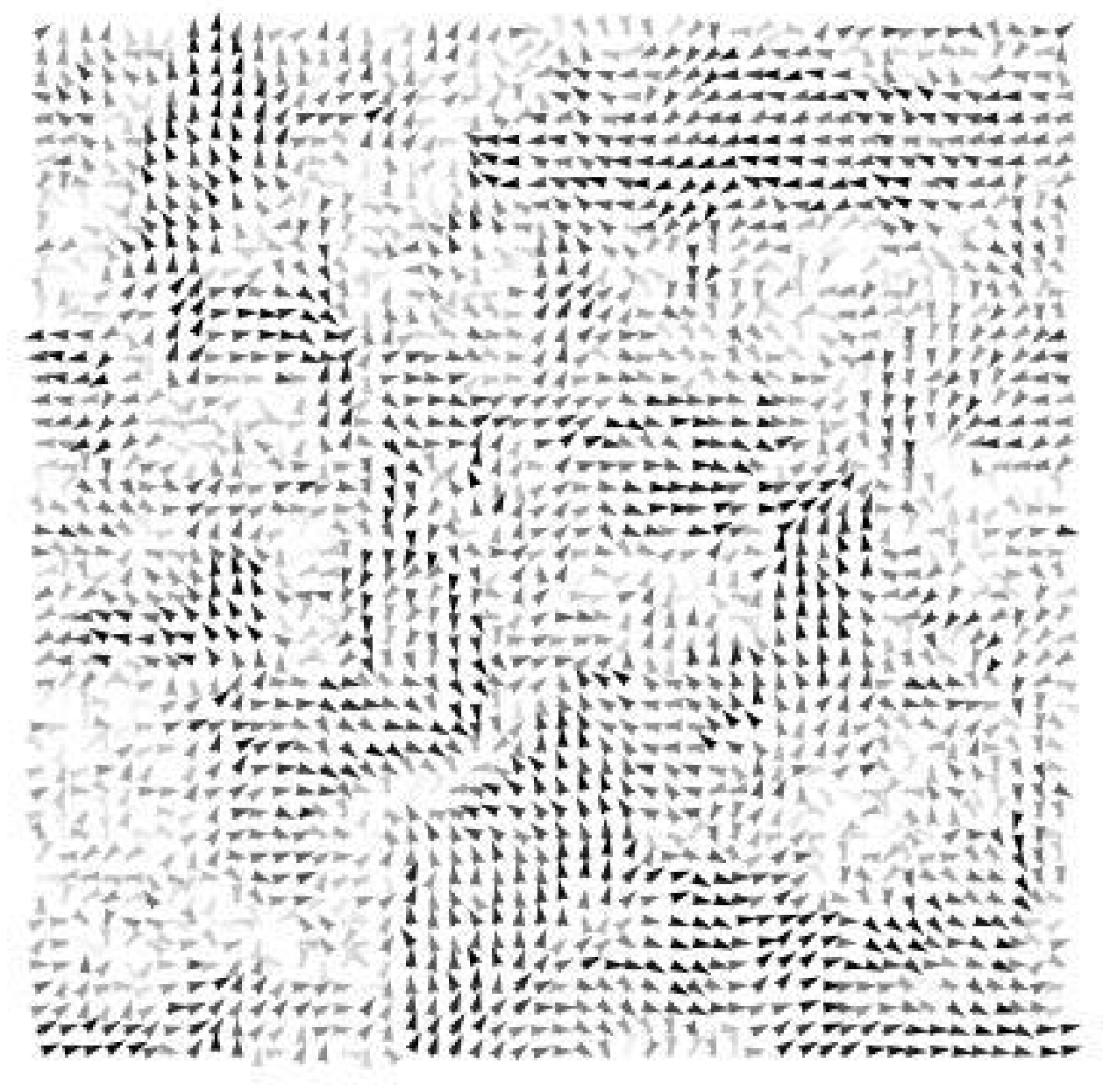}
\includegraphics[angle=-90,width=.23\textwidth]{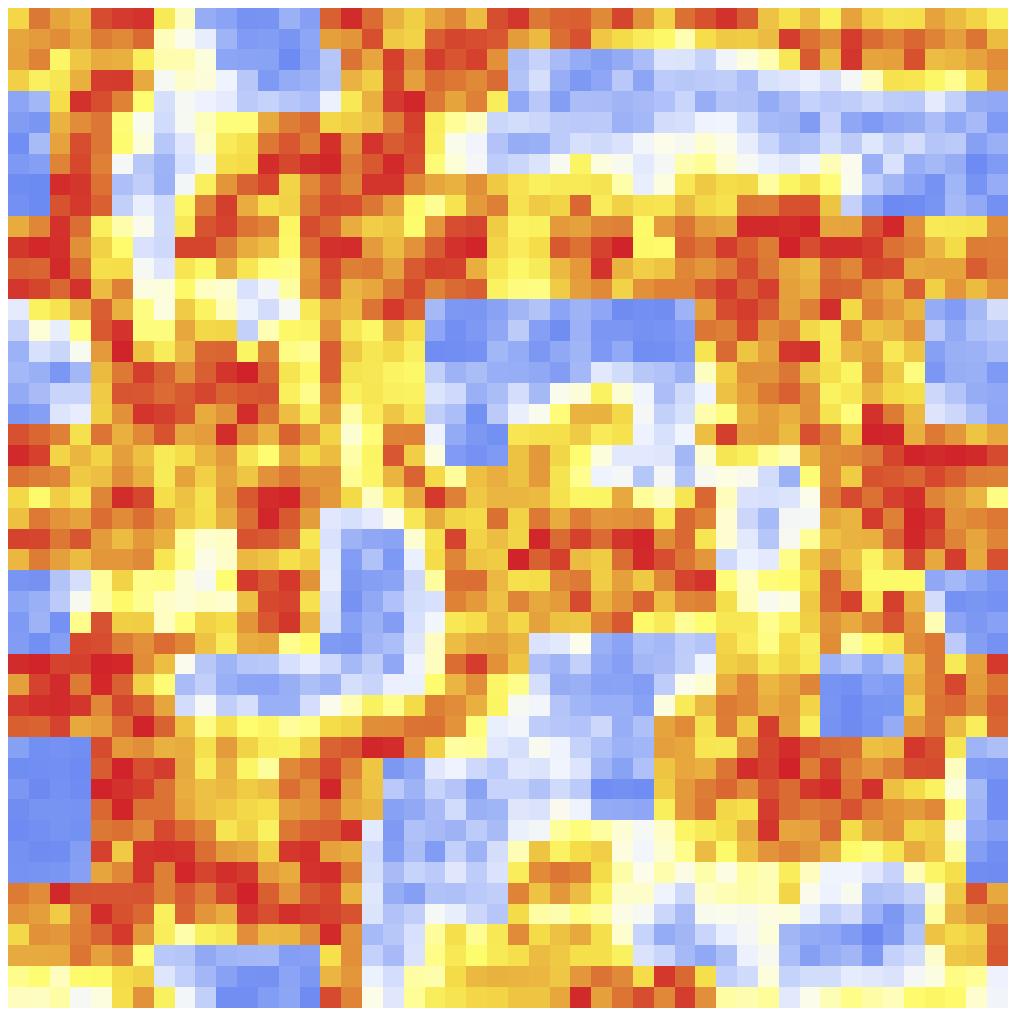}\\
\includegraphics[angle=-90,width=.23\textwidth]{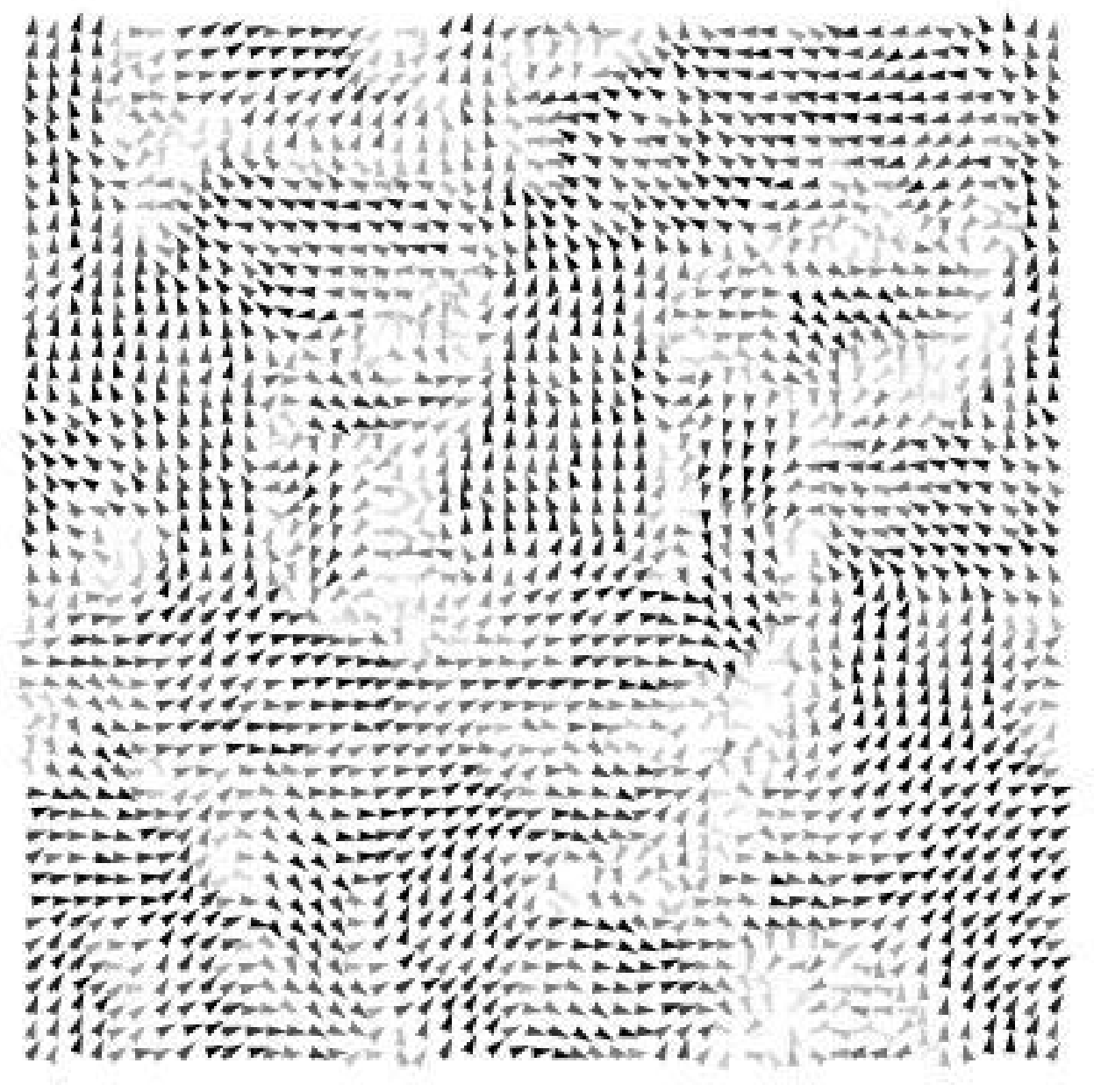}
\includegraphics[angle=-90,width=.23\textwidth]{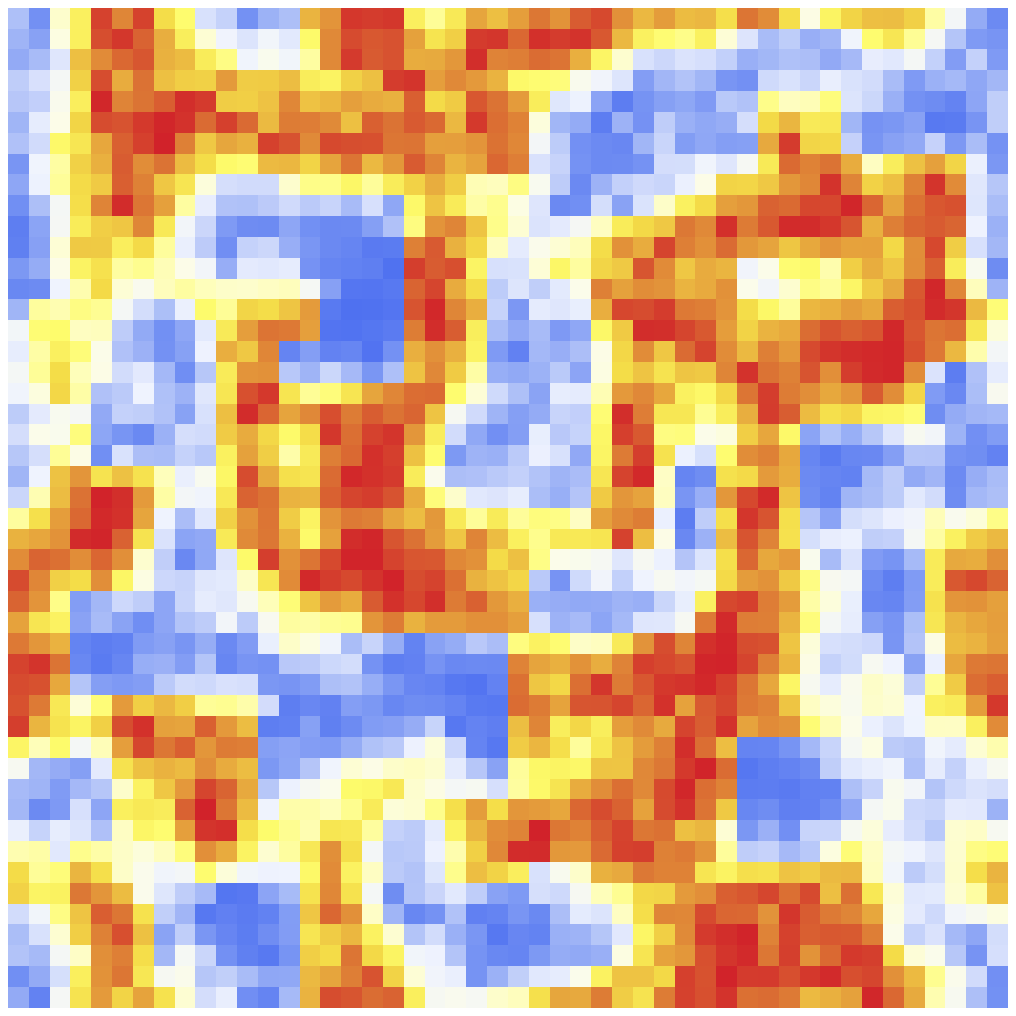}\\
\includegraphics[angle=-90,width=.23\textwidth]{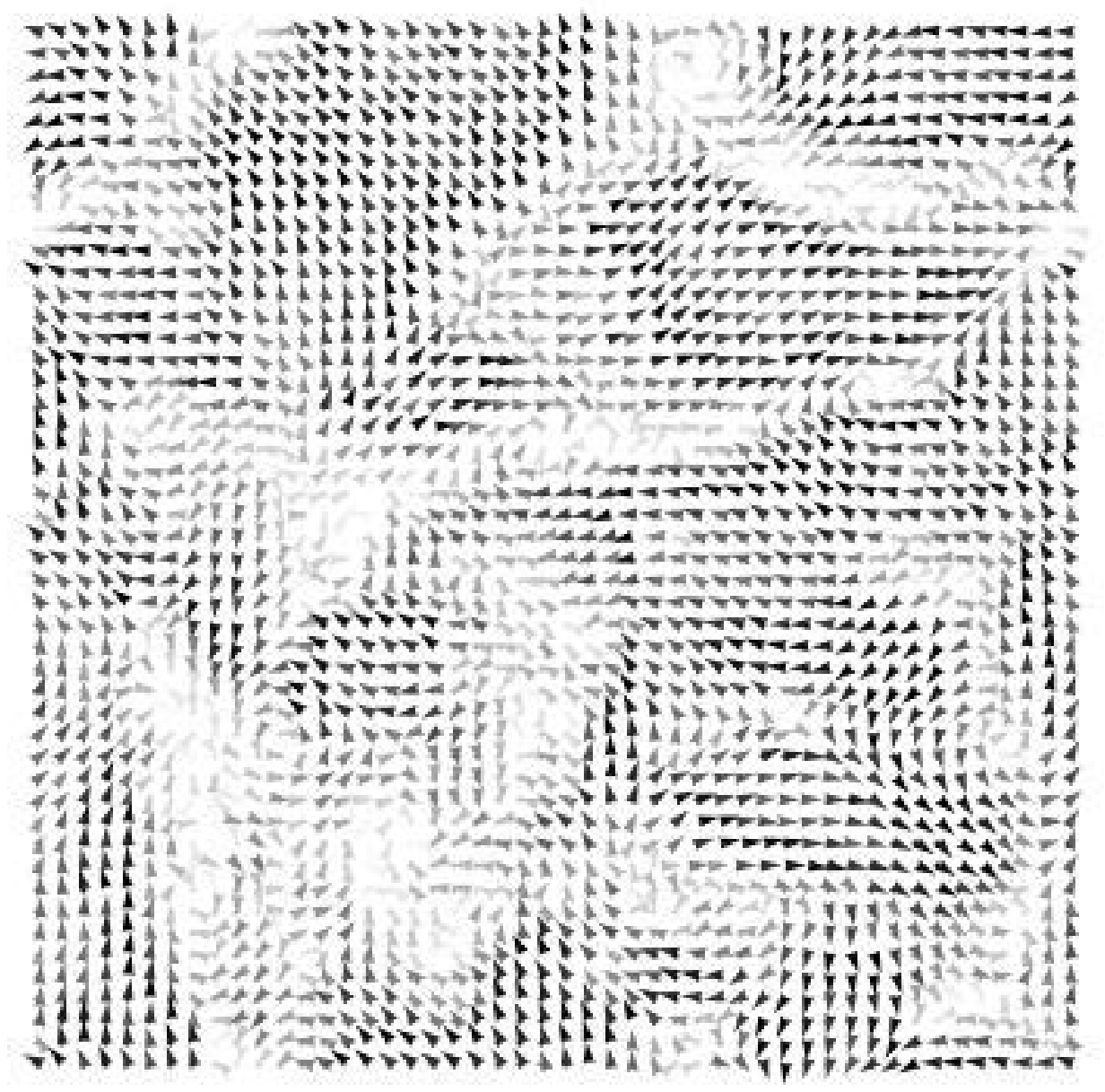}
\includegraphics[angle=-90,width=.23\textwidth]{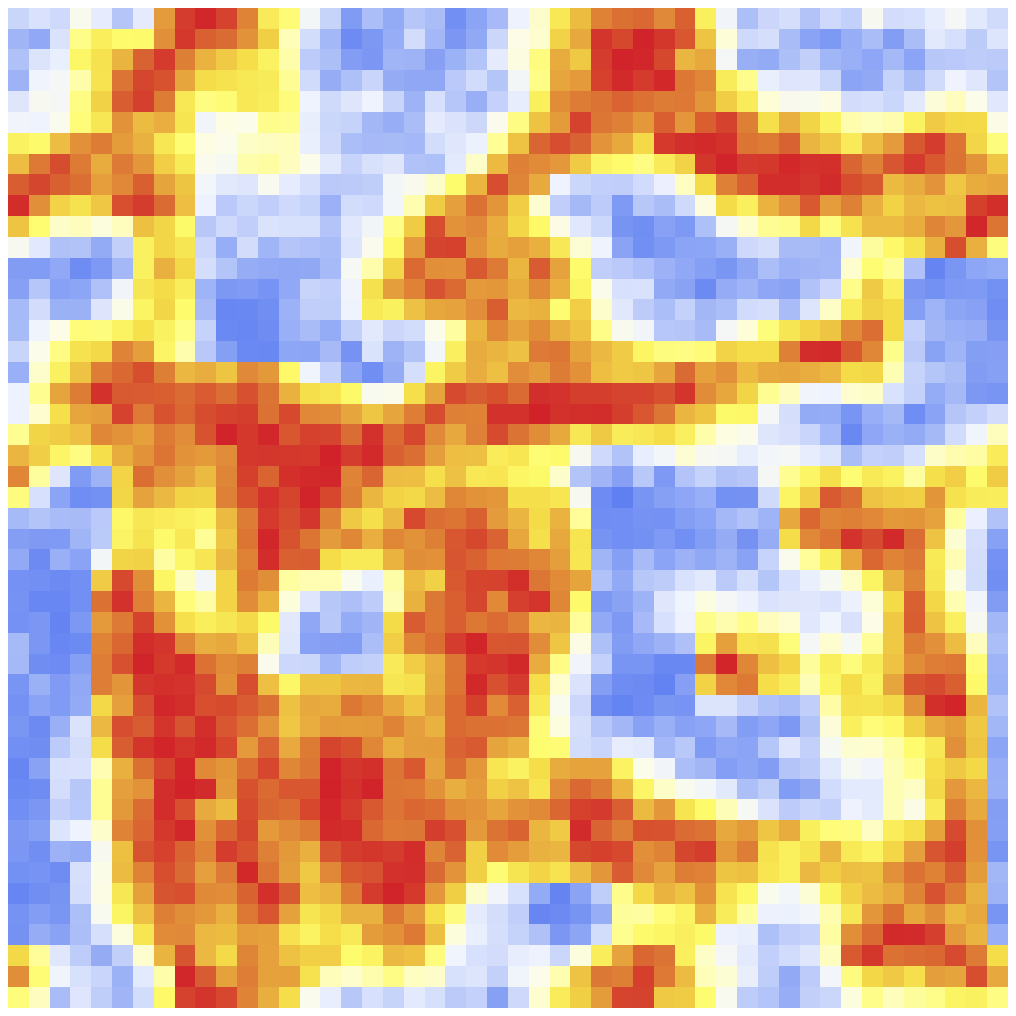}
\caption{Typical system configurations for the $\eta=5.5$ quench at $M_z=0.5$ and several temperatures below $T^*$. From up to down $T=0.4, 0.3, 0.2, 0.1$.}
\label{outconfMz.5} 
\end{center}
\end{figure} 

Figures \ref{inconfMz.5} and \ref{outconfMz.5} are typical configurations above and below $T^*$ respectively for which $M_z=0.5$, that is, in an equivalent macroscopic stage of the relaxation. The figures show several temperatures and again it is clear that the type of relaxation is different between the high and low temperature regions. Above $T^*$ the out-plane magnetization is mainly supported by uncorrelated  spins that remains with a non-zero out-plane projection. On the other hand, below $T^*$ the same magnetization is due to well-structured out-plane reminiscences that exactly corresponds with the defects of the cooperative in-plane ferromagnetic clusters. 
That is, for lower temperatures the depart from out-plane saturation is characterized by the movement of highly correlated domains of in-plane spins or, what is the same, by the slow movement of well-defined (out-plane) defects of these in-plane structures. 

\begin{figure}[hbt]
\begin{center}
\includegraphics[angle=-90,width=.23\textwidth]{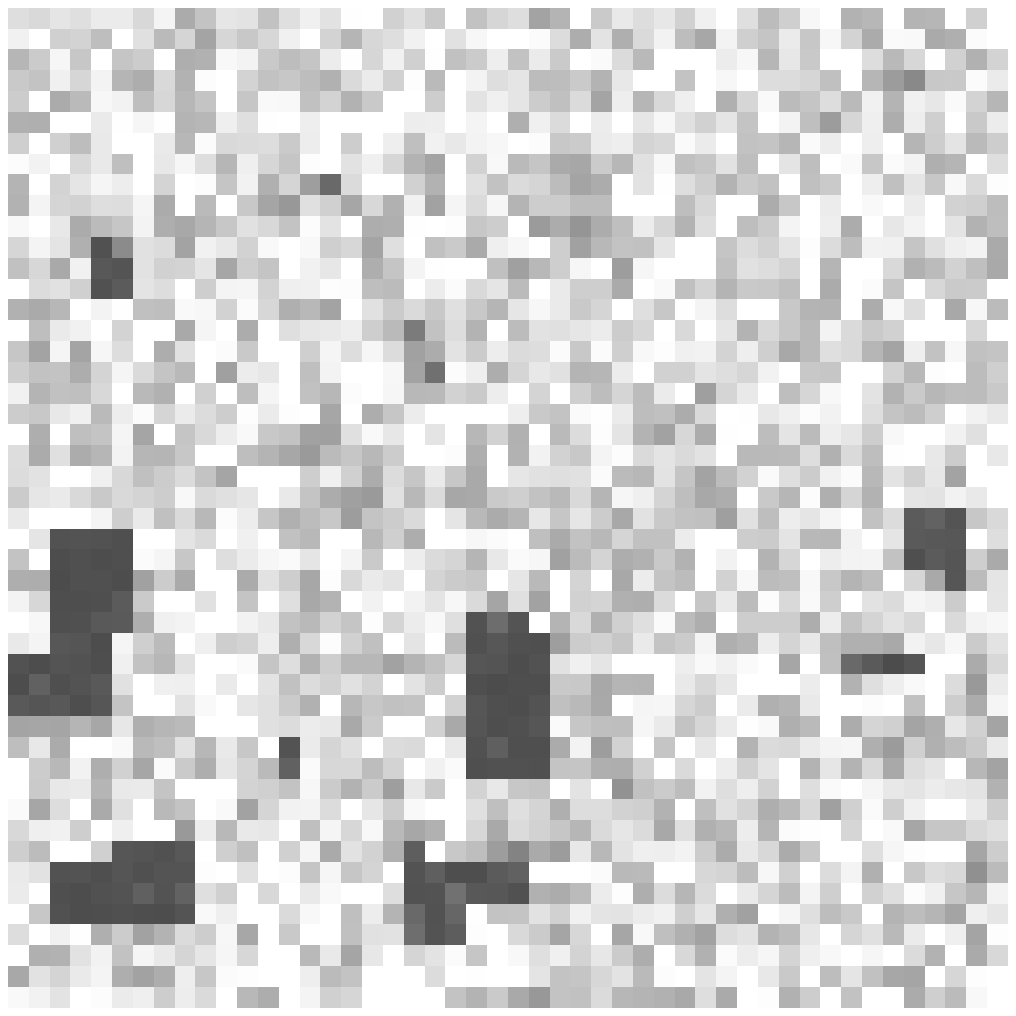}
\includegraphics[angle=-90,width=.23\textwidth]{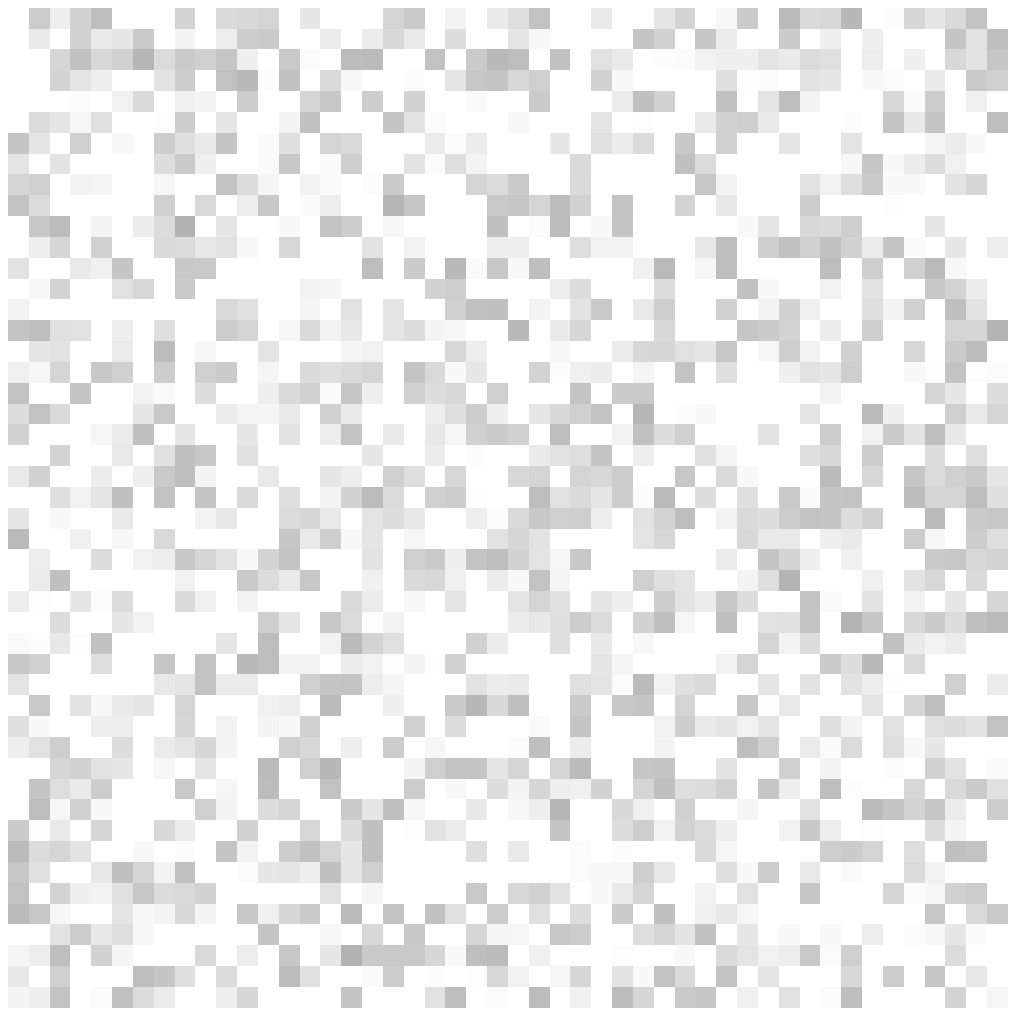}\\
\includegraphics[angle=-90,width=.23\textwidth]{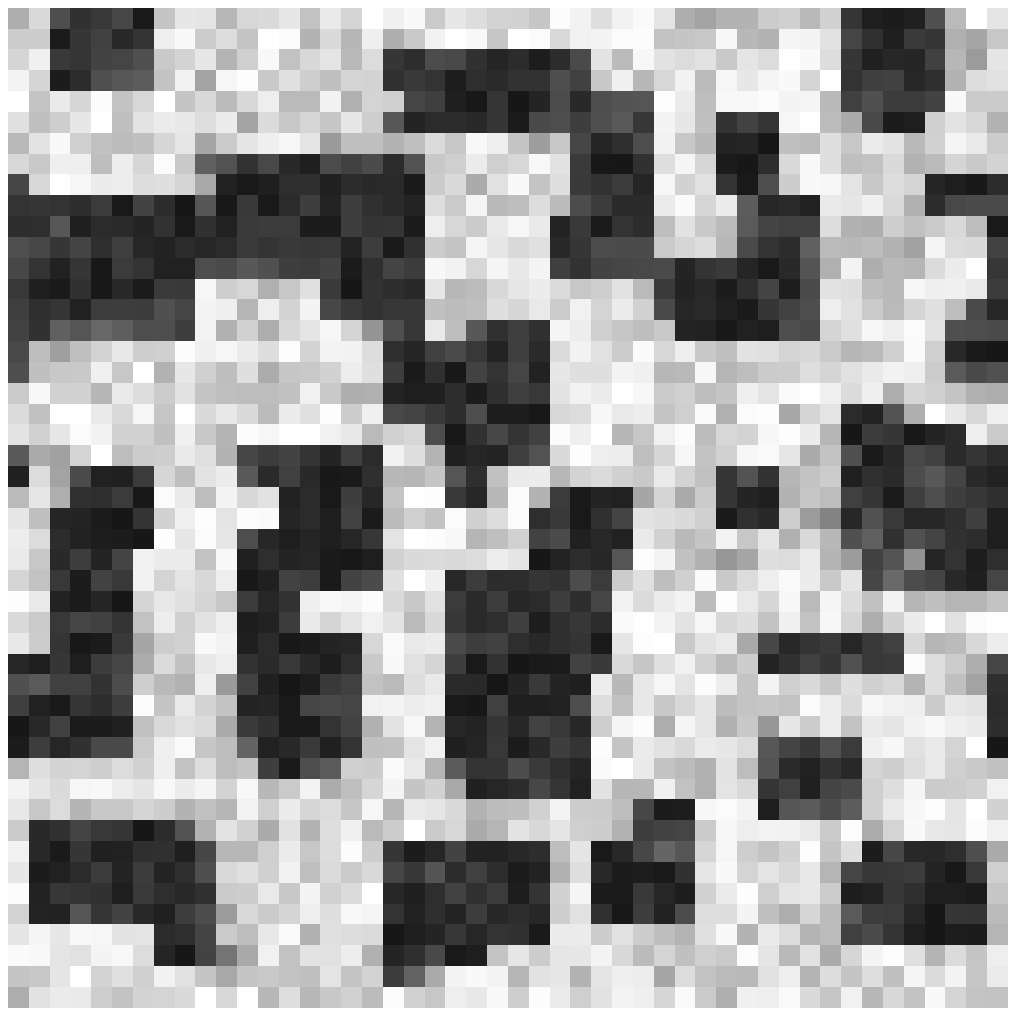}
\includegraphics[angle=-90,width=.23\textwidth]{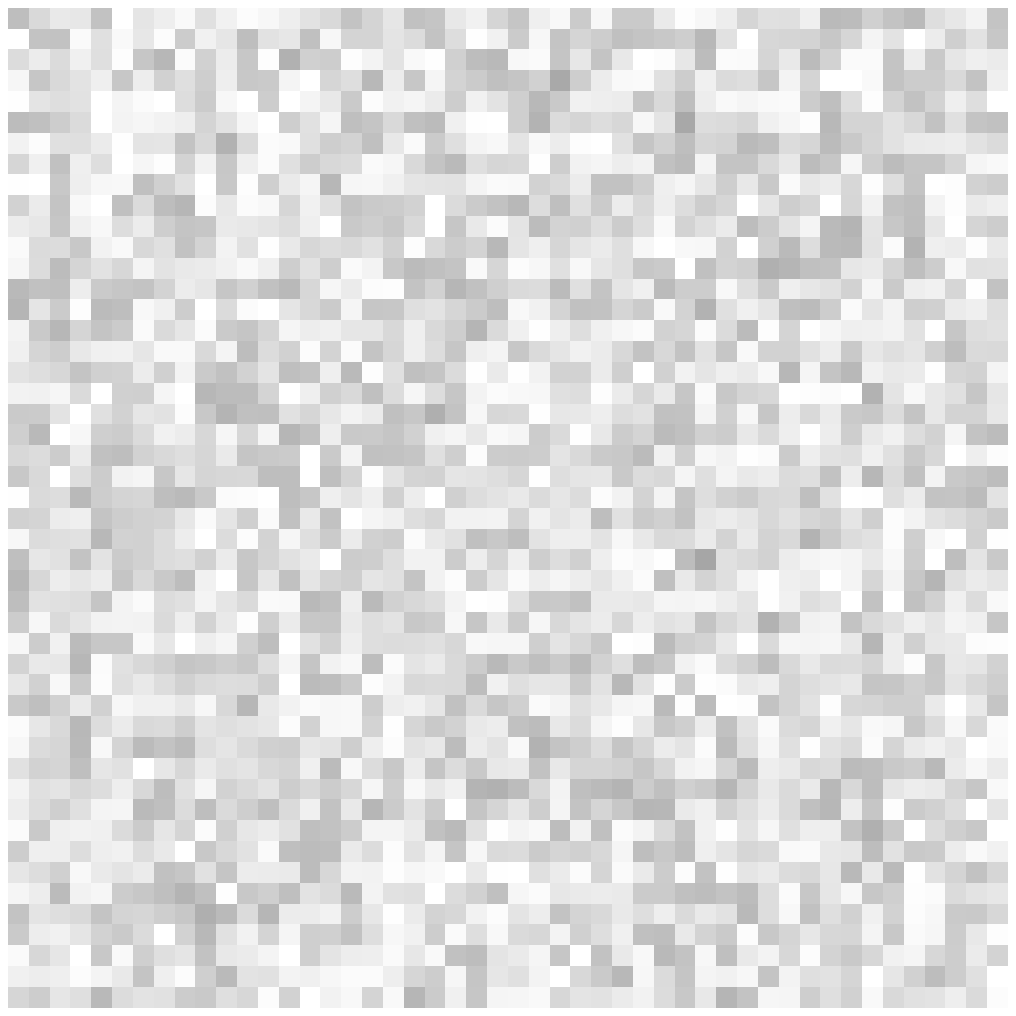}\\
\includegraphics[angle=-90,width=.23\textwidth]{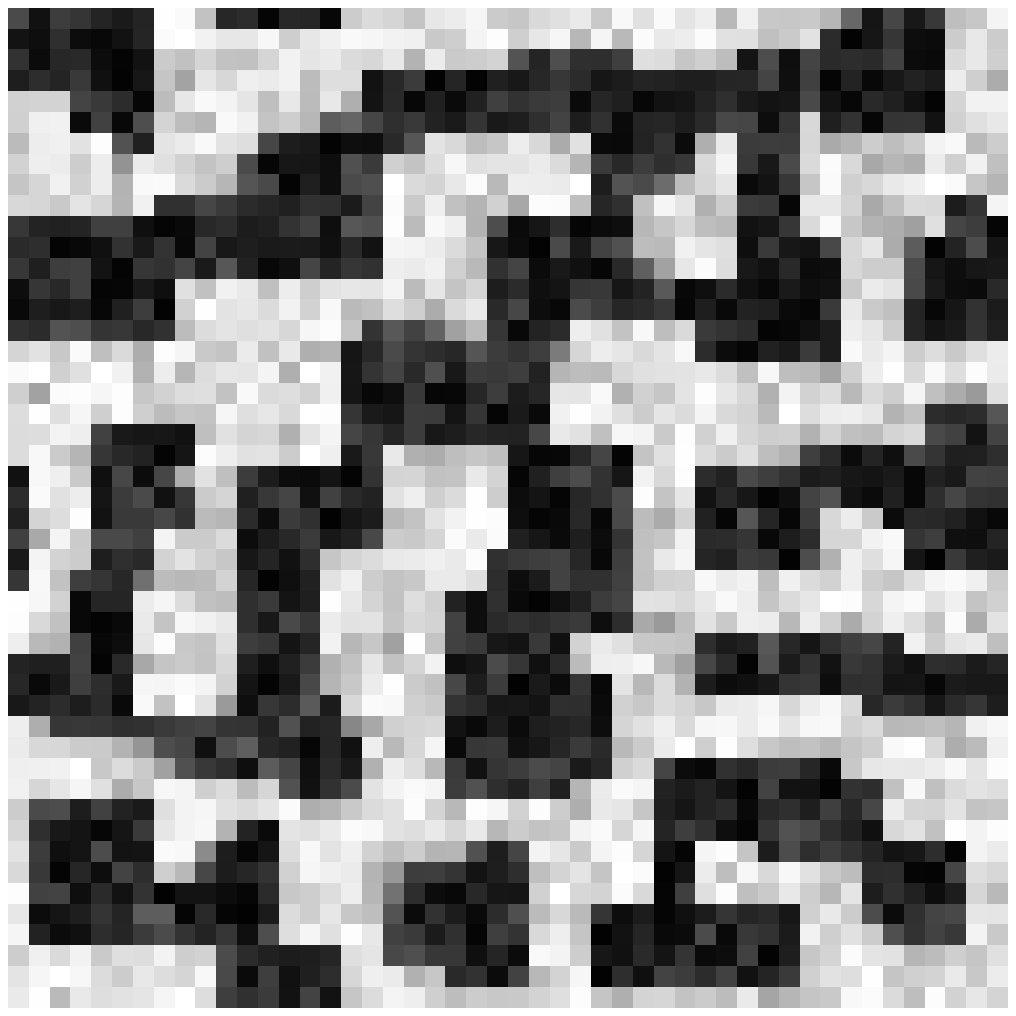}
\includegraphics[angle=-90,width=.23\textwidth]{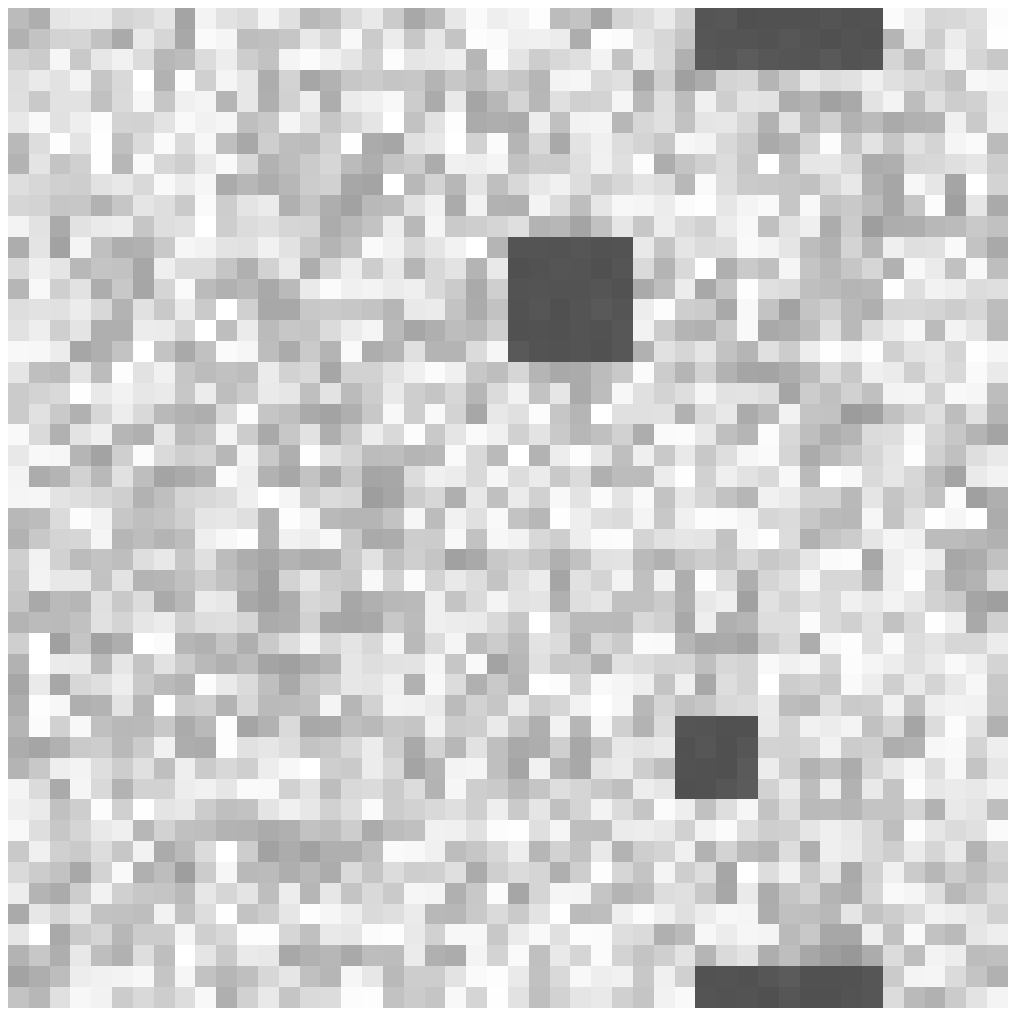}\\
\includegraphics[angle=-90,width=.23\textwidth]{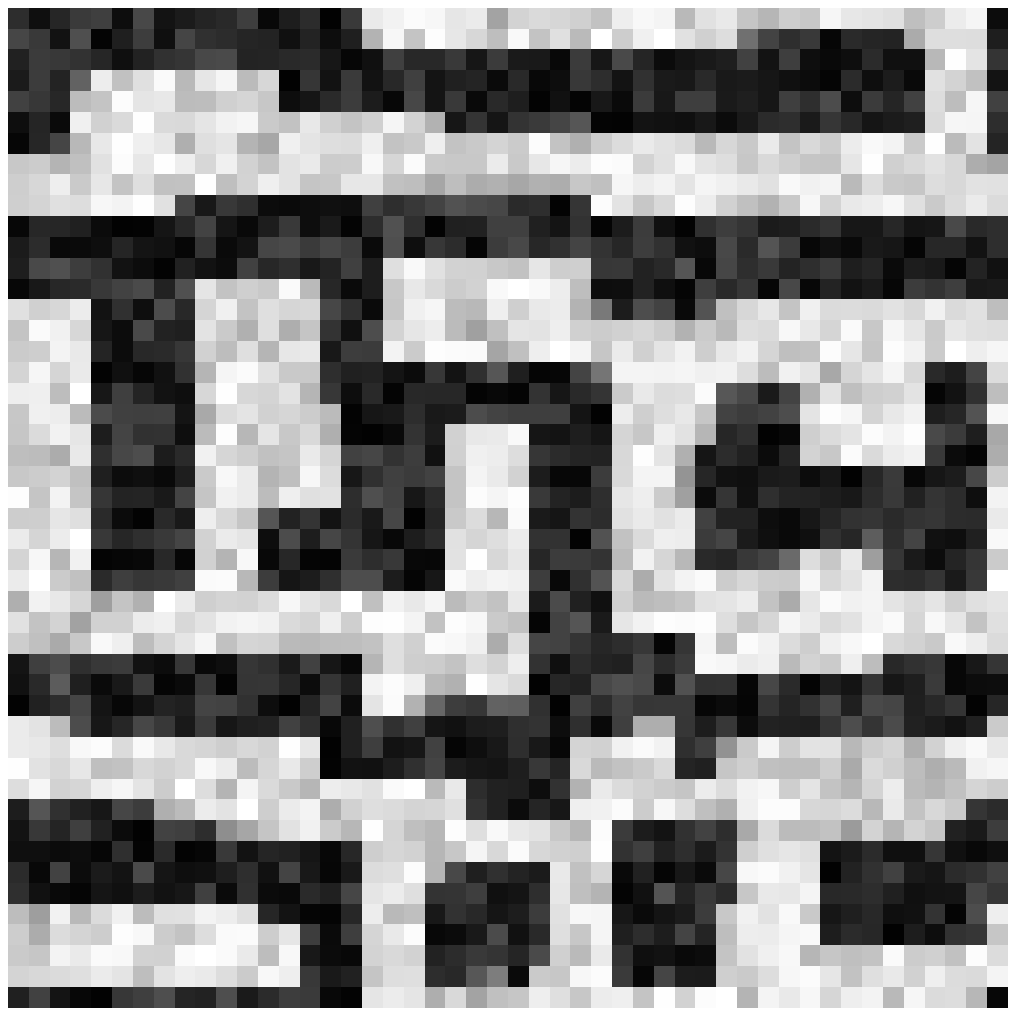}
\includegraphics[angle=-90,width=.23\textwidth]{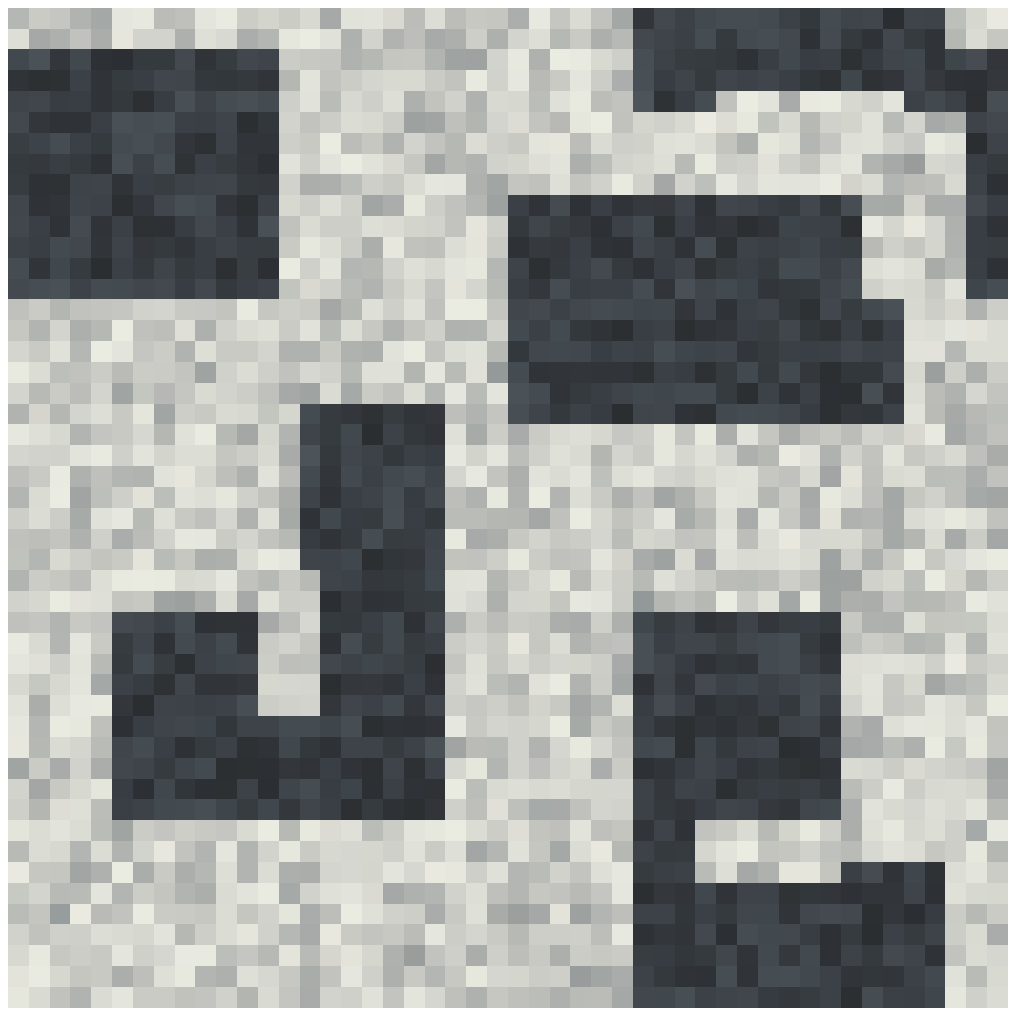}
\caption{Typical system out-plane spin projections ranging from $-1$ (black) to $1$ (white) for the $\eta=7.0$ quench at $T=0.8$ (left) and $T=0.4$ (right). From up to down correspondent times are $t=64, 256, 1024, 16385$.}
\label{outconfk7} 
\end{center}
\end{figure} 

Figure \ref{outconfk7} shows typical out-plane configurations for the quench at $\eta=7.0$. In this case in-plane views shows no order and no significant spin projections and therefore are uninteresting. As can be seen, the structure of the out-plane projections at $T=0.4$ are consistent with an increasing difficulty to reach the ground state. For higher temperatures the system is able to relax to a configuration in which the correct striped domains merges. On the other hand for $T=0.4$ the evolution is frustrated by an abnormal growth of the stripes width.

\subsection{Energy barriers distributions}
\label{ebd}

In previous works\cite{iglesias02,iglesias04} it has been shown that, in the range of validity of the $T\mathrm{ln}(t/\tau_0)$ scaling, the effective distribution of energy barriers ($f_b(E)$) contributing to the long time relaxation process can be obtained from the magnetic relaxation collapse simply by performing its logarithmic time derivative $dM(t)/d\mathrm{ln}(t)$. That is, studying the magnetic viscosity $S$. This distribution $f_b(E)$ represents a time independent distribution that would give rise to an identical relaxation curve. This quantity is calculated in this subsection from the numerical data.

Moreover, non-Arrhenius relaxation following equation (\ref{vftl}) can be seen as a simple rescaling of $T$. Thus, since the derivation of $S\sim f_b(E)$ does not involve temperature derivatives,\cite{iglesiasphd} the terms  remain unchanged and the relation between magnetic viscosity and energy barriers distribution still holds. So we are able to obtain the statistical information of the energy barriers above and below the dynamical transition temperature.

\begin{figure}[!htb]
\includegraphics[angle=-90,width=.45\textwidth]{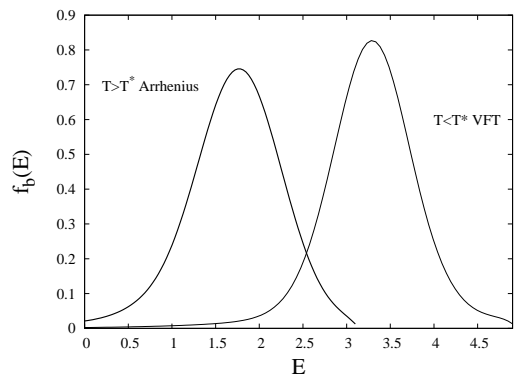}
\caption{Energy barriers distribution explored for the system with $\eta=5.5$ above and below $T^*$.}
\label{ebT} 
\end{figure} 

The results shown in figure \ref{ebT} for the two different dynamical regimes at $\eta=5.5$ confirms our expectations. Indeed, the slower non-Arrhenius relaxation involve higher energy barriers due to the cooperative nature of the domains movement.

\begin{figure}[!htb]
\includegraphics[angle=-90,width=.45\textwidth]{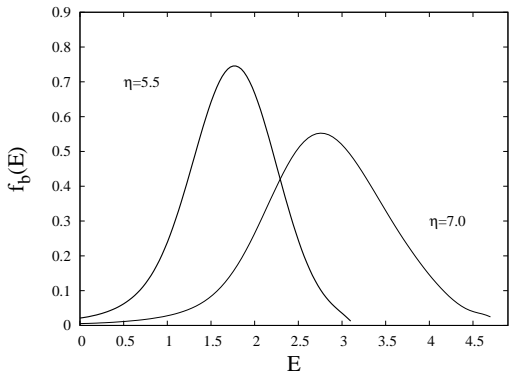}
\caption{Energy barriers distribution explored for the system at $T>T^*$ at both sides of the Spin Reorientation Transition.}
\label{ebk} 
\end{figure}

In the $\eta=7.0$ case some light can be spread on its anomalous behavior. If the energy barriers the system has to overcome are high enough, a Monte Carlo out-equilibrium dynamics freezes in some metastable configuration. So, it is worth to comparing the form of $f_b(E)$ between the quench to planar ferromagnetic phase and to the perpendicular striped region. Figure \ref{ebk} shows the result of analysis of energy barriers in the two regions in which the Arrhenius relation holds.

As can be seen, the distribution associated to the striped phase involves higher values of energy barriers than that of the planar ground state.  The shape of the distribution in the $\eta=7.0$ case is also wider suggesting a heterogeneous distribution of domains.     

This picture supports our believe that, once the system has formed striped structures, the energy barriers to be overcome are higher. Indeed they are so high that they can eventually freeze  below $T^*$.

\section{Conclusions}
\label{conc}

Through numerical Monte Carlo simulations we have studied the dynamical properties of a two dimensional Heisenberg model with dipolar interactions and perpendicular anisotropy. In particular, the analysis of the out-of-plane magnetization dynamics suggests the existence of a dynamical temperature transition $T^*$. Below this temperature the dynamics changes from an Arrhenius to a VFT type ($\eta=5.5$) or the system freezes ($\eta=7.0$). 

The interpretation of this anomalous slow down is understood as a result of the change from the single-spin elementary movement to a new mechanism based on the coherent movement of the oriented ground-state domains. The same explanation can be done in terms of the pinning of the defects merging in the frontiers of ground-state domains of different orientations. This explanation is confirmed by the analysis of typical configurations during the dynamics and by the shape of the energy barriers distributions obtained studying the magnetic viscosity of the model above and below $T^{*}$.


\section{Acknowledgments}
We gratefully acknowledge partial financial support from the Abdus Salam ICTP through grant {\em Net-61,
La\-tin\-a\-me\-ri\-can Network on Slow Dynamics in Complex Systems}. Calculation facilities kindly offered by the Bioinformatic's Group of the Center of Molecular Immunology were instrumental to the realization of this work.

\bibliography{heisenberg_v3.4}

\end{document}